\documentclass[12pt, oneside]{article}
\usepackage[letterpaper, left=1in, right=1in, top=1in, bottom=1in]{geometry}
\usepackage{setspace} 
\usepackage{lineno} 
\usepackage{siunitx}
\usepackage{amssymb}
\usepackage{amsmath}
\usepackage{dsfont}
\usepackage{mathptmx} 
\usepackage{graphicx}
\usepackage[labelfont=bf]{caption} 
\usepackage{float} 
\usepackage{booktabs}
\usepackage{hyperref}
\usepackage[ 
	style=authoryear,
	citestyle=authoryear,
	maxbibnames=6,
	giveninits=true,
    uniquename=init,
	backend=bibtex,
	url=false,
	doi=false,
	isbn=false,
	eprint=false
	]{biblatex}
\AtEveryBibitem{%
\clearfield{note}%
\clearfield{month}%
\clearfield{date}%
}
\usepackage{pdfpages}

\usepackage{tikz}
\usepackage{titlesec}
\titleformat{\section}[block]{\large\filcenter}{\thesection}{1em}{}
\titleformat{\subsection}[block]{\large\itshape\filcenter}{\thesubsection}{1em}{}
\titleformat{\subsubsection}[block]{\large\itshape\filcenter}{\thesubsubsection}{1em}{}
\titleformat{\paragraph}[runin]{\itshape}{\theparagraph}{1em}{}[. ]
\usepackage[normalem]{ulem}

\begin{document}



\title{{\vspace{-2cm}\LARGE\center Linking individual bioenergetics to ecosystem dynamics with integral projection models \\}} 
\author{}
\author{Willem Bonnaff\'e$^{1,\star}$, Martina Muraro$^{1}$, William Goulding$^{2}$, \\ Doug W. Smith$^{3}$, Dan R. Stahler$^{3}$, Peter Hudson$^{4}$, Hamish McCallum$^{2}$, \\ Sonya Clegg$^{1}$, Dan R. MacNulty$^{5}$, and Tim Coulson$^{1}$}
\date{}
\maketitle

\begin{center}
\vspace{-0.5cm}
1 Department of Biology, University of Oxford, Oxford, UK \\
2 Centre for Planetary Health and Food Security, Griffith University, Gold Coast, QLD, Australia \\
3 Yellowstone National Park, National Park Service, Mammoth Hot Springs, WY, USA \\
4 Department of Biology, and Huck Institutes of the Life Sciences, The Pennsylvania State University, University Park, PA, USA \\
5 Department of Wildland Resources and Ecology Center, Utah State University, Logan, UT, USA \\
$\star$ Corresponding author (willem.bonnaffe@biology.ox.ac.uk) 
\end{center}

\paragraph{Abstract} Linking individual level processes, such as survival and reproduction, to ecosystem dynamics is challenging due to the complexity of interactions within and across populations and species.
These interactions depend on the traits of the organisms, such as age, body size, and mass, and so are influenced by fluctuations in population structure. 
Structured population models, such as integral projection models (IPMs), have advanced our understanding of the dynamics that arise as a consequence of population structure in continuous traits.
Yet, these models have only been applied to relatively simple systems, featuring one or two species, only spanning a fraction of the full range of trophic levels observed in natural food webs.
In this work, we extend these models in four ways. 
First, we provide a general multi-species IPM where population structure can be included in primary producers, as well as primary and secondary consumers, thus achieving a wide range of ecological interactions. 
Second, we combine key individual-level processes to ecosystem functioning, by mapping fluxes of biomass between species to fluxes of biomass within individuals, using bioenergetic principles for allocation of biomass to maintenance, reproduction, or growth.
Third, we introduce a nutrient recycling loop through which decomposers break down organic matter into nutrients that enable plant growth.
Finally, we provide an efficient fitting algorithm to calibrate the model with time series of population sizes.
We showcase our approach by parameterising and fitting the model to time series of counts of elk, bison, wolves, and cougars in northern Yellowstone National Park to study the impact of predator extirpation and recovery.
Our model produces intriguing patterns, for instance, it predicts that the bison population decreases following the removal of predators, and recovers following predator recovery, suggesting a positive contribution of wolves and cougars to the recent increase of bison in northern Yellowstone.
It also reproduces known indirect effects of predators on woody deciduous vegetation regeneration. 
The flexibility of our approach makes it suitable for studying dynamics in a broad range of ecosystems, such as savannah, or marine systems, with application for population management and ecological forecasting.

\vspace{0.5cm}

%


\setlength{\parindent}{15pt}   

\begin{spacing}{1.1}

%


\newpage

\section*{\uppercase{Introduction}}

Understanding and predicting ecosystem dynamics has been a longstanding challenge in ecology (\cite{Lawton1999}). 
This is because it requires the integration of processes across multiple levels of organisation, from individuals to ecosystems (\cite{Turchin1999a}). 
Individuals of different species survive, grow, and reproduce by interacting with conspecifics, but also with individuals from other species, e.g. through competition and predation (\cite{Berryman2003}). 
The outcomes of these individual-level processes underpin changes at the population level (\cite{Turchin1999a, Berryman2002}), such as increases or decreases in the number of individuals, and shifts in the age, phenotypic, or genetic structure of the population. 
These shifts in population-level properties influence individual-level processes in other populations, such as those of competitors, prey, and predators, as well as biotic variables external to the population, hence introducing complex feedback effects of a population onto itself through a direct impact on its ecosystem (\cite{Berryman2003}). 
Trophic cascades are a particular example of such a sequence of effects, whereby the removal of a predator in an ecosystem can cause herbivore numbers to increase, which in turn can reduce vegetation biomass (\cite{Paine1980, Polis1996, Ripple2012, Passoni2024}). 
It is because of these multiple pathways and the ramifications of effects of interactions between species and their environment, and the multiplicity of the underlying mechanisms, that ecosystem-level dynamics are so challenging to model and predict (\cite{Lawton1999}). 

An additional aspect of nature that makes linking individual-level processes to ecosystem-level processes challenging is population structure, i.e. the fact that populations are made of heterogeneous groups of individuals with different trait values (\cite{Coulson2001}). 
These traits are often not fixed throughout the life of an individual (\cite{Childs2016a}). 
For instance, plastic traits, such as phenology, depend on the environment (\cite{Scheiner1993}), while morphological traits, such as body mass, change continuously as organisms grow (e.g. \cite{Bonnaffe2018}). 
This leads to so-called ontogenetic shifts in essential individual-level processes like food acquisition, survival, and reproduction (e.g. \cite{Specziar2009a, Mittelbach1998}). 
For example in fish, the diet of individuals often changes with body mass (\cite{Mittelbach1998, Specziar2009a}); small trout are essentially insectivorous while larger individuals become almost exclusively piscivorous (\cite{LabeeLund1992}). 
In large terrestrial herbivores, e.g. elk (\textit{Cervus canadensis}) and bison (\textit{Bison bison}), survival increases with ontogeny, as smaller individuals can be more easily captured by predators, compared to larger individuals (\cite{Ruth2019, Becker2009a}), although this also depends on age due to senescence. 
Reproductive capacity also varies throughout growth; many organisms, humans for example, do not have the capacity to reproduce in early stages of life. 
Therefore, the way individuals interact with other components of their ecosystem and allocate resources to either growth or reproduction depends on their age and growth stage (\cite{Coulson2001, Ozgul2010, Childs2016a}). 

Population structure thus introduces transience in ecological interactions, which makes predicting consequences for population dynamics of interacting species a difficult exercise. 
This problem is at the heart of structured population modelling (SPMs), a mathematical framework that explores the dynamics of systems that feature populations divided into groups of individuals with similar properties (\cite{Caswell1988}). 
These models enable the depiction of simplified life histories, that is, the transition of groups of individuals between different life stages (\cite{Caswell1988}). 
A key distinction within SPMs lies in the mathematical nature, discrete or continuous, of the structuring variables. 
Matrix population models divide populations into discrete categories, for instance by considering a juvenile class and an adult class (\cite{Caswell1988}). 
However, this discretisation is often arbitrary, as most phenotypic traits, such as body mass, are continuous (\cite{Ellner2006}). 

Integral projection models (IPMs) describe how distributions of continuous traits change between two consecutive time steps (\cite{Ellner2006}).
This enables the representation of individual-level variation in traits present in natural populations without having to track explicitly each individual throughout its life. 
IPMs thus capture fine-scale variation in the demographic state of populations, in terms of the age and growth stages that are present, and consequently, in the interactions within and between populations. 
Individual-based models (IBMs) consider each individual as an agent, and though they can thereby introduce more realistic mechanisms in the system, they do so at the expense of mathematical tractability of many functions, and computational efficiency, due to the need for multiple repeat simulations to average over the stochasticity of individual events.
This makes IPMs a suitable alternative to link individual- and population-level processes.
To define an IPM, one needs to specify key processes, such as survival, reproduction, and growth, and define how interactions, either competitive or predatory, vary with ontogeny, to observe the type of population dynamics that emerge. 
Though they portray the distribution of continuous traits, in practice they are usually approximated as large matrices to numerically solve the integrals (\cite{Ellner2006}).
These models have been developed to study the consequences of multiple types of interactions on population dynamics, such as intra-specific density-dependence in Soay sheep (\cite{Simmonds2015}), intra-specific predation in perch and pike (\cite{Claessen2000, Ohlberger2019}), inter-specific competition in perennial plant communities and in corals (\cite{Adler2010, Chu2015, Kayal2018}), intraguild predation in guppies and killifish (\cite{Bassar2023}), and herbivory and predation across two and three trophic levels in Yellowstone (\cite{Lachish2020, Passoni2024}).
Beyond species interactions, some studies have also made attempts at incorporating the wider ecosystem into structured population models.
This is the case of the Demographic Ecosystem modelling framework, which tracks the dynamics of populations of plants structured by phenotype and of their biogeochemical environment (e.g. \cite{Moorcroft2001, Longo2019}).

In spite of these advances linking specific aspects of individual-, population-, and ecosystem-level processes, SPMs still fall short of achieving a minimum viable depiction of an ecosystem, that is one that integrates the dynamics of plants, herbivores, predators, decomposers, nutrients, and decomposing organic matter. 
Studies using IPMs usually consider up to three species with structured populations spanning one or two trophic levels at most, often with a single type of interaction, either competitive or predatory. 
In reality, population structure interplays with intra- and inter-specific interactions across all trophic levels, not just the ones that are under examination. 
For instance, in Yellowstone, aspen and willow experience browsing by elk up until a certain height, after which they largely escape browsing (\cite{Hobbs2024, Brice2024}). 
So, the ontogeny of woody vegetation conditions the capacity of elk to access this particular food source. 
This calls for a framework that encompasses phenotype structure across multiple trophic levels.
To our knowledge, no IPM has introduced the same level of population structure in primary producers, primary consumers, and secondary consumers, and integrated multiple types of interactions (i.e. intra- and inter-specific competition and predation) by having at least two species at each trophic level.
Achieving this would hence require at least six structured species (i.e. 2 species $\times$ 3 trophic levels).
In addition, one of the most fundamental aspects of ecosystem functioning is nutrient recycling.
The dynamics of structured populations should thus influence nutrient recycling to some extent in order for the model to capture ecosystem dynamics.
Finally, even if such a model existed, it would ultimately need to be grounded in a real system in order for us to have any degree of confidence that it is producing plausible dynamics.
We argue that no study to date achieves this, due to a lack of a suitable methodology for fitting large models to time series data, though pioneering work has made strides in this direction (\cite{Ghosh2012, Elderd2016, White2016}).
Overall, despite significant potential, the suitability of IPMs for studying the links between individual-level processes and ecosystem dynamics thus remains limited.

We extend SPMs in four ways to achieve a more complete and realistic representation of ecosystem dynamics. 
First, by enabling the inclusion of multiple species with structured populations in three trophic levels and the associated structured interactions, both competitive and predatory.
Second, by using a simple bioenergetic model, we link ecosystem fluxes to the ontogeny of individuals, connecting ecosystem-level processes to individual-level processes.
Third, by introducing nutrient recycling, an essential aspect of ecosystem functioning which introduces indirect effects between primary producers and secondary consumers.
Finally, by providing an algorithm to ground the model in a real system through fitting time series of count data.
To showcase our approach, we demonstrate how this framework can be used to study ecosystem dynamics in northern Yellowstone National Park.
We are not aware of studies that have attempted to fit IPMs with more than one population to time series data (\cite{Ghosh2012, Elderd2016, Gonzalez2016, White2016}).
Our aim in this case study was to assess whether the more complicated models that we consider here, featuring bioenergetics, multiple species, and nutrient cycling, could be parametrised and fitted to longitudinal datasets in a way that produces plausible dynamics evidenced by (i) the existence of an equilibrium state in our model of northern Yellowstone where all species coexist, (ii) biomass distributions and densities that are consistent with field observations, and (iii) a response of the model to predator extirpation and reintroduction broadly consistent with changes in vegetation and herbivores recorded in northern Yellowstone over the past 30 years.
Beyond northern Yellowstone, the flexibility of our approach, which can be parameterised easily to include new species, makes it suitable for a wide range of systems and questions.

\section*{\uppercase{Model description}}

In this section, we build a general bioenergetic ecosystem integral projection model (Eco-IPM) to study how interactions between structured species within and across trophic levels influence ecosystem dynamics. 

At its core, our framework models how the biomass distribution, namely the distribution of the body mass of individuals, of each species in the ecosystem changes from one time step to the next, through survival, reproduction, and growth (Fig. 1, A).
We also consider unstructured compartments of the ecosystem. 
These compartments are not structured by phenotype, either because they are composed of multiple taxa, or because they are abiotic.
We consider organic matter, nutrients (e.g. phosphates and nitrogen), decomposers (e.g. bacteria, fungi, and invertebrates), and unstructured plants, focussing on grass.
Reproduction and growth are limited by the mass of resources that individuals are able to consume in other compartments of the ecosystem (Fig. 1, C, bottom panel), while survival depends on their lifespan and is conditioned by their ability to avoid being consumed by predators, or herbivores in the case of plants (Fig. 1, C, top panel). 
The mass of resources consumed by the different populations is obtained from four different sources: (i) nutrient absorption, (ii) herbivory, (iii) scavenging, or (iv) predation (Fig. 1, B).
In order for the main types of interactions typically present in an ecosystem to take place, we consider at least six structured species in addition to the unstructured compartments: two competing primary producers (plants), two competing primary consumers (herbivores), and two competing secondary consumers (predators) (Fig. 1, B, square nodes).
For the sake of simplicity, we only consider indirect competition caused by sharing resources.
Future implementation of direct competition into the framework could contrast results obtained to test the importance of direct competition for the observed demographic patterns.
Finally, we implement a nutrient recycling loop controlled by the decomposer community, which introduces feedbacks between secondary consumers and primary producers (Fig. 1, B, brown arrows).

The following subsections provide the mathematical details necessary to build this general Eco-IPM.
We first define the population dynamics of structured plants and animals.
We then describe how individuals within populations survive, acquire resources from their abiotic and biotic environment, and allocate consumed resources to maintenance, reproduction, and growth. 
Finally, we define nutrient recycling by computing the dynamics of organic matter, nutrients, and decomposers, thereby producing the full ecosystem model.
We carefully designed the main equation of the model such that it describes population dynamics of both structured plants and animals. 
Differences between species are achieved through differences in the parameter values.
This is to keep the model as general as possible and facilitate the subsequent inclusion of additional species and application to other ecosystems than the ones considered in the case study.

\newpage
\thispagestyle{empty}
\begin{figure}[H]
  \centering  
  \includegraphics[width=\textwidth, page=1]{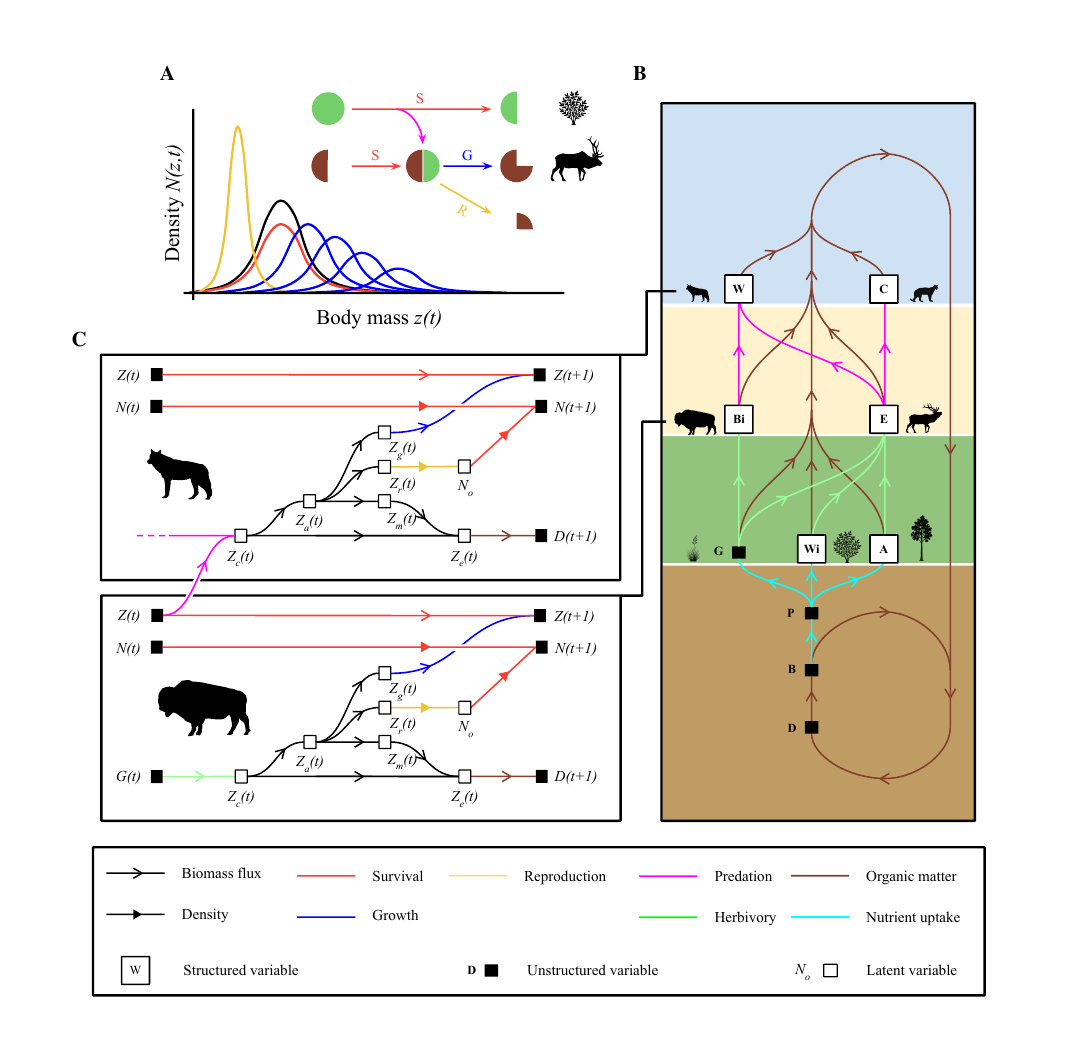}
  \caption{
    \textbf{Conceptual diagram of the bioenergetic ecosystem integral projection model.}
    Panel A shows how the population body mass distribution (in black) changes through applying the survival (S, in red), growth (G, in blue), and reproduction (R, in gold) kernels.
    Panel B describes the fluxes of biomass between the different compartments of the ecosystem, which are either unstructured (organic matter (D), decomposers (B), nutrients (P), and grass (G)) or structured (willow (Wi), aspen (A), elk (E), bison (Bi), wolves (W), and cougars (C)).
    Structured compartments are represented by a biomass distribution, rather than by a single point estimate of the total amount of biomass.
    Panel C shows how ingoing biomass fluxes are captured ($Z_c$), assimilated ($Z_a$), and allocated to metabolic maintenance ($Z_m$), reproduction ($Z_r$), and growth ($Z_g$).
    Growth feeds directly into the total population biomass ($Z(t)$), while the number of recruits ($N_o$), increases the total population density.
    Predators are modelled as an additional source of mortality whereby they capture a given amount of biomass from other populations.
    Fluxes of biomass that occur through scavenging of biomass of dead bison and elk are not shown for clarity. 
    Those would be shown as a fraction of organic matter coming out of elk and bison into a carcass compartment, a fraction of which would be captured by predators.
    Figure and icons created by Willem Bonnaff\'e.
  }
  \label{fig:conceptual-diagram}
\end{figure}

\subsection*{Parameters}

The parameters in the equations are designated by greek letters and are divided into functional groups for each species. 
Parameters that relate to the ontogeny of survival, reproduction, and resource capture are designated by the letter $\beta$, those that control the proportion of the different food sources acquired by $\lambda$, those relating to consumption and assimilation by $\alpha$, $\gamma$, and $\rho$, those linked to probability density functions of recruitment and growth by $\mu$ and $\sigma$, and those that determine metabolic costs and reproductive investment by $\delta$ and $\nu$, respectively. 
Parameters that describe ecosystem fluxes between unstructured compartments, by $\phi$, and rates, by $\rho$.
Let 
\[
    \Omega^{(i)} = \left\{\beta^{(i)}, \alpha^{(i)}, \gamma^{(i)}, \rho^{(i)}, \mu^{(i)}, \sigma^{(i)}, \delta^{(i)}, \nu^{(i)} \right\}
\]
denote the set of parameters for structured species $i$ and $\Omega^{(0)} = \left\{\phi^{(0)}, \rho^{(0)}, \gamma^{(0)} \right\}$ the set of parameters for unstructured species and abiotic compartments.
The superscript $X^{(0)}$ is used to emphasise the difference between unstructured and structured variables.
All parameter notations, along with their unit and meaning, are presented in Table 1.

\setstretch{1.0}
\begin{table}
\centering
\caption{
\textbf{Summary of parameter notations and meaning.}
The table is divided into unstructured compartments, i.e. decomposers and perennial grasses, and structured components, i.e. woody deciduous plants, herbivores, and carnivores.
The superscript $(i)$ indicates that there is one parameter for each structured species.
}
\label{tab:model_comparison}
\resizebox{\linewidth}{!}{
\begin{tabular}{llllll}
\hline
\\
\textbf{Parameter} & \textbf{Unit} & \textbf{Meaning} & \textbf{Parameter} & \textbf{Unit} & \textbf{Meaning} \\
\\
\midrule
\\
\multicolumn{6}{l}{\textit{Unstructured compartments}} \\
\addlinespace
\multicolumn{3}{l}{\textbf{Decomposers}} & \multicolumn{3}{l}{\textbf{Perennial grasses}} \\
$\phi_\text{BD}$                      & yr$^{-1}$                 & Mortality rate                 & $\phi_\text{GD}$                    & yr$^{-1}$                 & Mortality rate            \\
$\phi_\text{DB}$                      & --                        & Prop. organic matter available & $\phi_\text{PG}$                    & --                        & Prop. nutrients available \\
$\rho_\text{B}$                       & kg kg$^{-1}$ yr$^{-1}$    & Consumption rate               & $\rho_\text{G}$                     & kg kg$^{-1}$ yr$^{-1}$    & Consumption rate          \\
$\phi_\text{DP}$                      & --                        & Prop. waste-product            & $\gamma_3^{(\text{G})}$             & kg kg$^{-1}$              & Nutrient efficiency       \\
                                      &                           &                                &                                     &                           &                           \\
\addlinespace
\multicolumn{6}{l}{\textit{Structured compartments}} \\
\addlinespace
\multicolumn{3}{l}{\textbf{Survival}} & \multicolumn{3}{l}{\textbf{Diet}} \\
$\beta_{s,n,0}^{(i)}$                 & yr$^{-1}$                 & Maximum survival rate          & $\lambda_\text{1}^{(i)}$            & --                        & Prop. uptaking nutrient   \\
$\beta_{s,n,1}^{(i)}$                 & kg                        & Inflection point               & $\lambda_\text{2}^{(i)}$            & --                        & Prop. grazing             \\
$\beta_{s,n,2}^{(i)}$                 & kg$^{-1}$                 & Slope                          & $\lambda_\text{3}^{(i)}$            & --                        & Prop. scavenging          \\
                                      &                           &                                & $\lambda_\text{4}^{(i)}$            & --                        & Prop. consuming kills     \\
\multicolumn{6}{l}{\textbf{Growth and consumption}} \\
$\alpha_1^{(i)}$                      & --                        & Prop. assimilation             & \multicolumn{3}{l}{\textbf{Metabolic Costs}} \\
$\alpha_2^{(i)}$                      & --                        & Prop. conversion               & $\delta^{(i)}$                      & kg kg$^{-0.75}$ yr$^{-1}$ & Metabolic costs           \\
$\gamma_1^{(i)}$                      & kg yr$^{-1}$              & P.c. consumption               &                                     &                           &                           \\
$\gamma_2^{(i)}$                      & --                        & Prop. consumable biomass       & \multicolumn{3}{l}{\textbf{Nutrient uptake}} \\
$\sigma_g^{(i)}$                      & kg                        & S.d. of growth                 & $\beta_{d,u,0}^{(i)}$               & indiv$^{-1}$ yr$^{-1}$    & P.c. uptake rate          \\
                                      &                           &                                &                                     &                           &                           \\
\multicolumn{3}{l}{\textbf{Reproduction and Recruitment}} & \multicolumn{3}{l}{\textbf{Grazing}} \\
$\beta_{b,0}^{(i)}$                   & yr$^{-1}$                 & Maximum reproduction rate      & $\beta_{d,h,0}^{(i)}$               & indiv$^{-1}$ yr$^{-1}$    & P.c. grazing rate         \\
$\beta_{b,1}^{(i)}$                   & kg                        & Weight at reproduction onset   &                                     &                           &                           \\
$\beta_{b,2}^{(i)}$                   & kg$^{-1}$                 & Slope                          & \multicolumn{3}{l}{\textbf{Intra- and interspecific mortality}} \\
$\nu_o^{(i)}$                         & --                        & Prop. biomass for reproduction & $\beta_{d,p,0}^{(i,j)}$             & indiv$^{-1}$ yr$^{-1}$    & Max. probability          \\
$\mu_o^{(i)}$                         & kg                        & Mean of offspring weight       & $\beta_{d,p,1}^{(i,j)}$             & kg                        & Inflection point          \\
$\sigma_o^{(i)}$                      & kg                        & S.d. of offspring weight       & $\beta_{d,p,2}^{(i,j)}$             & kg$^{-1}$                 & Slope                     \\
                                      &                           &                                &                                     &                           &                           \\
\\
\hline
\end{tabular}
}
\end{table}
\normalsize

\subsection*{Population dynamics}

We use IPMs to represent the dynamics of structured populations of woody deciduous plants, herbivores, and carnivores. 
IPMs are integrodifference equations, namely equations that require solving an integral to compute how a continuous distribution changes between two discrete time steps (\cite{Ellner2006}).
These models are typically used to describe the dynamics of distributions of phenotypes, such as body size or mass, allowing the operator to incorporate realistic demographic processes into population dynamics, such as ontogeny and sexual maturation.

\textcite{Smallegange2017} initially incorporated bioenergetics into IPMs following the principles of dynamic energy budget models for resource conversion into energy and allocation to growth or reproduction (\cite{Kooijman2010}).
\textcite{Lachish2020} extended this framework by considering body mass of elk as the focal phenotype, which explicitly defines how the biomass of grass consumed is assimilated, converted into energy, and allocated to physiological processes such as somatic maintenance, reproduction, and growth, and by tracking grass biomass fluctuations due to consumption by elk. 
We extend this framework by providing a more general bioenergetic IPM framework where the same general equation describes the dynamics of biomass distributions of plants, herbivores, and carnivores: 
\begin{equation}\label{eq:ipm}
N^{(i)}(z, t + 1) = \int_{S_z} \left \{ f_r(z | z') p_{b}(z') n_{o}(z') + f_g(z | z') \right \} p_s(z') N^{(i)}(z', t) dz',
\end{equation}
where $N^{(i)}(z, t) \in \mathbb{R}^{+}$ is the density of individuals of species $i$ with phenotype $z \in \mathbb{R}^{+}$, defined as the body mass, at time $t$.
$S_z$ is the support of the phenotypic trait values, i.e. $[0,+\infty[$.
$f_r(z|z')$ and $f_g(z|z')$ are the probability density functions of the recruitment and growth, respectively, to phenotype $z$ given an initial phenotype $z'$.
The terms $p_{b}(z')$ and $p_s(z')$ correspond to the probability of reproducing and surviving, respectively, from time $t$ to $t+1$.
The term $n_{o}(z')$ is the average number of offspring produced by a single individual with phenotype $z'$.
This construction assumes a post-breeding census where changes in the biomass distribution occur first through survival, then through growth and reproduction.

\subsection*{Survival}

\paragraph{Overall survival}
Following a post-breeding census, individuals must survive before they can reproduce or grow (Fig. 1, red, pink and green lines). 
We consider two sources of mortality, extrinsic, where individuals are killed by consumers (Fig. 1, pink and green lines), i.e. herbivores and predators, and intrinsic, where individuals die from other causes of death that do not involve consumers (Fig. 1, red lines). 
Individuals need to survive both intrinsic and extrinsic causes of mortality, 
\begin{equation}
p_s^{(i)}(z) = p_{s,n}^{(i)}(z) p_{s,p}^{(i)}(z),
\end{equation}
where $p_{s}^{(i)} \in [0,1]$ is the probability that an individual from species $i$ with body mass $z$ survives until the next year, and $p_{s,n}^{(i)}$ and $p_{s,p}^{(i)} \in [0,1]$ are the probabilities that it survives intrinsic causes of death and consumption by herbivores or predators, respectively.
This construction is mathematically equivalent to that in a similar study by \textcite{Bassar2023}, but differs in its biological meaning as it depends on body mass, and not body length.
Because we consider body mass as the phenotype, this allows us to determine the mass of resources acquired by consumers, which cannot be calculated directly by using body length.
In addition, we acknowledge that this dichotomy between intrinsic and consumer-related mortality ignores additional sources of mortality, such as disease or hunting, which can be substantial.
Finally, we also note that this construction assumes that intrinsic mortality is independent on extrinsic mortality, which may not always be the case in real systems. 
For instance, animals that have a high intrinsic mortality because they are affected by a disease may be more susceptible to predation.
These are simplifying assumptions intended to keep our framework as modular as possible.

\paragraph{Survival to intrinsic causes of death}
As the survival of individuals often increases throughout growth (e.g. \cite{White2015, Lachish2020, Smith2020, White2024}), we model individual intrinsic survival as a sigmoid function of body mass,
\begin{equation}
p_{s,n}^{(i)}(z) = \beta_{s,n,0}^{(i)} \cdot \mathrm{sigmoid} \left\{\left(z - \beta_{s,n,1}^{(i)}\right) \beta_{s,n,2}^{(i)} \right\},
\end{equation}
where $\beta_{s,n,0}^{(i)} \in [0,1]$ is the maximum, or asymptotic, survival probability, $\beta_{s,n,1}^{(i)} \in \mathbb{R}$ is the body mass at maximum increase in log-odds of surviving per unit of phenotype, also known as the inflection point, $\beta_{s,n,2}^{(i)} \in \mathbb{R}$ is the slope, and $\mathrm{sigmoid}\{x\} = 1/(1+\exp\{-x\})$.
This function controls how the survival odds of individuals change as they grow in the absence of consumers. 

\paragraph{Survival to consumers}
We assume that the probability to survive encounters with consumers should increase with the body mass, as larger individuals are often more resistant to consumers, and decrease with the number of consumers.
To reflect this in our model we define the probability of surviving consumption by $J$ different species of consumers following a Binomial model, i.e. a sequence of Bernoulli trials (see supplementary section S4 for derivation),
\begin{equation}
p_{s,p}^{(i)}(z) = \prod_{j=1}^{J} p_{s,p}^{(i,j)}(z),
\end{equation}
where $p_{s,p}^{(i,j)}(z)$ denotes the probability that an individual of species $i$ with body mass $z$ avoids consumption by species $j$. 
Following the same logic, this quantity itself depends on the number of encounters with individual consumers of that species following 
\begin{equation}
p_{s,p}^{(i,j)}(z) = \prod_{k=1}^{N^{(j)}(t)}p_{s,p}^{(i,j,k)}(z),
\end{equation}
where $p_{s,p}^{(i,j,k)}(z)$ is the probability that an individual of species $i$ survives an encounter with a single consumer $k$ of species $j$, and $N^{(j)}(t)$ is the number of individuals of species $j$ present at time $t$.

Predators and herbivores tend to target younger individuals, for instance, elk in northern Yellowstone primarily browse on smaller aspen trees (\cite{Brice2024}). 
So, we introduce a decrease in the probability of consumption with increasing body mass, using the sigmoid function
\begin{equation}
p_{d,p}^{(i,j,k)}(z) = 1 - p_{s,p}^{(i,j,k)}(z) = \beta_{d,p,0}^{(i,j)} \cdot \mathrm{sigmoid} \left\{\left(z - \beta_{d,p,1}^{(i,j)}\right) \beta_{d,p,2}^{(i,j)}\right\},
\end{equation}
where $\beta_{d,p,0}^{(i,j)}$ denotes the maximum probability of death of an individual of species $i$ with phenotype $z$ from consumption by an individual of species $j$, $\beta_{d,p,1}^{(i,j)}$ the body mass at which individuals become increasingly resistant to consumption, and $\beta_{d,p,2}^{(i,j)}$ the slope, or steepness, of this transition.

\paragraph{Partitioning mortality of structured species} 
One of the key contributions of this work is that we couple the demography of interacting structured populations by explicitly tracking the biomass that is transferred from a population to the consumer populations.
To determine the mass and origin of resources that consumers feed on, we compute the proportion of deaths in each population attributable to the different consumer species.
To do this, we partition the probability of death into additive contributions of the different consumers.
We make use of the relations $p_s = 1 - p_d$ and $\sum_j p_{d,p}^{(i,j)}(z) = \sum_k p_{d,p}^{(i,k)}(z)$, such that their ratio is equal to 1, as follows
\begin{equation}\begin{aligned}
p_{s}^{(i)}(z) &= p_{s,n}^{(i)}(z) p_{s,p}^{(i)}(z), \\
\Leftrightarrow p_{d}^{(i)}(z) &= 1 - p_{s,n}^{(i)}(z) p_{s,p}^{(i)}(z), \\
&= p_{d,n}^{(i)}(z) + p_{s,n}^{(i)}(z) p_{d,p}^{(i)}(z), \\
&= p_{d,n}^{(i)}(z) + p_{s,n}^{(i)}(z) p_{d,p}^{(i)}(z) \left( \frac{\sum_{j=1}^{I} p_{d,p}^{(i,j)}(z)}{\sum_{k=1}^{I} p_{d,p}^{(i,k)}(z)} \right), \\
&= p_{d,n}^{(i)}(z) + p_{s,n}^{(i)}(z) p_{d,p}^{(i)}(z) \sum_{j=1}^{I} \frac{p_{d,p}^{(i,j)}(z)}{\sum_{k=1}^{I} p_{d,p}^{(i,k)}(z)}. \\
\end{aligned}\end{equation}
In essence, this formula divides mortality into additive terms, each of which is proportional to the probability of consumption by a particular consumer species relative to the total consumption probability.
This allows us to compute the corresponding amount of biomass that each consumer population removes from the consumed population,
\begin{equation}\begin{aligned}\label{eq:partition-predation}
Z_{d}^{(i)}(t) &= \int z' p_d^{(i)}(z') N^{(i)}(z', t) dz', \\
&= \int z' \left( p_{d,n}^{(i)}(z') + p_{s,n}^{(i)}(z') p_{d,p}^{(i)}(z') \sum_{j=1}^{I} \frac{p_{d,p}^{(i,j)}(z')}{\sum_{k=1}^{I} p_{d,p}^{(i,k)}(z')} \right) N^{(i)}(z', t) dz', \\
&= \int z' p_{d,n}^{(i)}(z') N^{(i)}(z', t) dz' + \sum_{j=1}^{I} \int z' p_{s,n}^{(i)}(z') p_{d,p}^{(i)}(z') \frac{p_{d,p}^{(i,j)}(z')}{\sum_{k=1}^{I} p_{d,p}^{(i,k)}(z')} N^{(i)}(z', t) dz', \\
&= Z_{d,n}^{(i)}(t) + \sum_{j=1}^{I} Z_{d,p}^{(i,j)}(t), \phantom{\int} \\
\end{aligned}\end{equation}
where $Z_d^{(i)}(t) \in \mathbb{R}^{+}$ is the total biomass of species $i$ that dies between $t$ and $t+1$, $Z_{d,n}^{(i,j)}(t)$ is the biomass of species $i$ that died from intrinsic causes, and $Z_{d,p}^{(i,j)}(t)$ corresponds to the biomass of individuals removed by species $j$.

\paragraph{Partitioning mortality of unstructured resources}
Some species in our system, such as woody deciduous plants, consume nutrients, noted $P(t)$, while herbivores consume different species of grasses, noted $G(t)$, both of which are resources that are not considered structured by phenotype in our framework.
To partition the mass of unstructured resources into different sources of removal, we apply the same procedure that we used for partitioning structured species biomass. 
The difference is that we can simplify the integrals in the previous equations, Eq. \ref{eq:partition-predation}, given that nutrient mass and grass biomass are independent of the phenotype, $z$.
Taking the example of grass, we arrive at
\begin{equation}\begin{aligned}
G_{d}^{(0)}(t) &= p_{d,n}^{(0)} \cdot G^{(0)}(t) + \sum_{j=1}^{J}  p_{s,n}^{(0)} \cdot p_{d,h}^{(0)} \cdot \frac{p_{d,h}^{(0,j)}}{\sum_{k=1}^{I} p_{d,h}^{(0,k)}} G^{(0)}(t), \\
&= G_{d,n}^{(0)}(t) + \sum_{j=1}^{J} G_{d,h}^{(0,j)}(t), \\
\end{aligned}\end{equation}
where $G_d^{(0)}(t)$ is the biomass of dead grass, $G_{d,n}^{(0)}$ and $G_{d,h}^{(0,j)}$ are the biomass of grass that died from intrinsic causes, and from herbivory by species $j$, respectively. 
We use the superscript convention $X^{(0)}$ here to emphasise that grass is an unstructured ecosystem compartment separate from structured species which are superscripted with the index $i$.
The probability of survival to herbivory is modelled as for structured species,
\begin{equation}
p_{s,h}^{(0)} = \prod_{j=1}^{J} \prod_{k=1}^{N^{(j)}(t)}p_{s,h}^{(0,j,k)}.
\end{equation}

The only difference with structured species is that $p_{s,h}^{(0,j,k)}$, the probability of surviving an encounter with an individual of species $j$, is a constant which does not depend on the phenotype,
\begin{equation}
p_{s,h}^{(0,j,k)} = 1 - \beta^{(0, j)}_{d,h,0},
\end{equation}
where $\beta^{(0, j)}_{d,h,0}$ is the probability of death following an encounter with a single herbivore.
The same equations are used to determine nutrient removal by the different deciduous plant species.

\subsection*{Acquisition and assimilation of resources}

Partitioning mortality into additive contributions of consumers is the key to compute the fluxes of biomass between populations.
This allows us to compute exactly how much biomass is removed from a population by a consumer population.
However, for animals, removing biomass from a population by killing individuals does not guarantee that they manage to acquire this biomass.
This depends on their capacity to defend these resources from other species, and sometimes from other groups of individuals from the same species.
So, we split biomass uptake into two steps: (i) acquisition, which corresponds to the utilisation of the biomass of kills for animals, and (ii) assimilation, which portrays internal processes of nutrient extraction and conversion into the individuals' own biomass.

\paragraph{Acquisition of resources}
Up to this point we have considered that individuals can directly capture resources from their biotic and abiotic environment, which for plants is achieved through root uptake of nutrients, and through herbivory and predation for herbivores and carnivores, respectively. 
Some species may also indirectly obtain resources from other species through scavenging by tapping into the pool of biomass of individuals that died from intrinsic causes.
Altogether this provides four general means of acquiring resources: (i) nutrient uptake, (ii) herbivory, (iii) scavenging, and (iv) predation. 
We compute the total acquired resource mass (Fig. 1, C, $Z_c(t)$) by a species by summing its capacity to tap into these four pools, 
\begin{equation}\begin{aligned}
Z_c^{(i)}(t) &= \lambda_1^{(i)} P_{d,u}^{(0,i)}(t) + \lambda_2^{(i)} G_{d,h}^{(0,i)}(t) +  \lambda_3^{(i)} \sum_{j=1}^{J} Z_{d,n}^{(j)}(t) + \lambda_4^{(i)} \sum_{j=1}^{J} Z_{d,p}^{(j,i)}(t), \phantom{\int} \\
\end{aligned}\end{equation}
where $Z_c^{(i)}$ is the total acquired mass of resources by the $i^{th}$ species, $\lambda_1^{(i)} \in [0,1]$ is the proportion of time spent taking up nutrients, $\lambda_2^{(i)}$ grazing, $\lambda_3^{(i)}$  scavenging, and $\lambda_4^{(i)}$ occupying kills.
$P_{d,u}^{(0,i)}$ is the mass of nutrients taken up and $G_{d,h}^{(0,i)}$ is the biomass of grass grazed.
These times are subjected to constraints: individuals cannot employ all four means of sustenance simultaneously which implies $\sum_k \lambda_{k}^{(i)} \leq 1$.
In addition, species have to share biomass with other species that may also allocate time to scavenging, such that $\sum_i \lambda_3^{(i)} \leq 1$; this ensures that total biomass acquired through scavenging across all species does not exceed the available biomass of carcasses.

\paragraph{Assimilation of acquired resources} 
Once the amount of acquired resources has been determined for each species, they consume it at a given rate.
Consumption is limited in two ways. 
First, the total mass of acquired resource is computed at the scale of the entire population, hence individual consumption should decrease with increasing population density.
Second, not all elements of the acquired resources can be consumed, e.g. not all parts of carcasses are equally available to all species.
These constraints effectively prevent over-consumption, i.e. consuming a mass of resources that exceeds the acquired mass. 
After consumption, resources are assimilated more or less efficiently. 
For herbivores, this depends on the nature of the material ingested, for instance, plant fibers are harder to break down and require specific adaptations.
As a consequence, a fraction of resources consumed is simply released without being assimilated.
Furthermore, the conversion of assimilated compounds into compounds that can be metabolised and mobilised for structural biomass production, reproductive tissue generation, or metabolic maintenance, is achieved at a loss.
This introduces constraints on assimilation and conversion of the consumed resources.
To implement those limitations, we choose a saturating model, equivalent to a Michaelis-Menten model for a fixed population size, to compute the total mass of resources consumed and successfully assimilated by a single individual in a year (Fig. 1, C, $Z_a(t)$),
\begin{equation}
Z_a^{(i)}(z,t) = \frac{\alpha_1^{(i)} \alpha_2^{(i)} \gamma_{1}^{(i)} \gamma_{2}^{(i)} Z_c^{(i)}(t) }{\gamma_{1}^{(i)} N^{(i)}(t) + \gamma_{2}^{(i)} Z_c^{(i)}(t)},
\end{equation}
where $Z_a^{(i)}(z,t)$ is the yearly mass of resources consumed and assimilated by an individual of species $i$ with phenotype $z$, $\alpha_1^{(i)} \in [0,1]$ is the assimilation efficiency, $\alpha_2^{(i)}$ is the conversion efficiency of assimilated resources into biomass of species $i$, which depends on the quality of the food consumed (e.g. nutrient content), $\gamma_{1}^{(i)} \in \mathbb{R}^+$ is the per-capita resource %
uptake rate of species $i$, $\gamma_{2}^{(i)} \in [0,1]$ is the fraction of the total acquired resources that can be consumed by species $i$, and $Z_c^{(i)}(t)$ denotes the total mass of acquired resources.
We note that we made an important simplifying assumption here, in that the consumption rate $\gamma_{1}^{(i)}$, and by extension $Z_a^{(i)}(z,t)$, are independent on the phenotype, $z$.
In reality, individuals with larger body masses tend to consume more resources than smaller ones.
We still write the consumption as a function of the phenotype to keep the framework as general as possible in the subsequent equations and facilitate future implementations that account for the dependence of consumption on body mass.
Overall, $Z_a^{(i)}(z,t)$ is effectively the net biomass gain for an individual of species $i$ with phenotype $z$ (Fig. 1, C).

\subsection*{Maintenance}
Assimilated biomass then becomes available for fulfilling physiological functions. 
We consider three main functions: (i) somatic maintenance, (ii) gametogenesis (i.e. production of reproductive cells), and (iii) somatic growth (Fig. 1, C, $Z_m(t)$, $Z_r(t)$, and $Z_g(t)$). 
Somatic maintenance costs increase with body mass.
Metabolic theory provides a well-supported mathematical formulation for this relationship (\cite{Brown2004a}),
\begin{equation}
Z_{m}^{(i)}(z, t) = \delta^{(i)} z^{3/4},
\end{equation}
where $Z_{m}^{(i)}(z, t)$ is the biomass allocated to maintenance and $\delta^{(i)} \in \mathbb{R}^{+}$ is a taxa-specific scaling coefficient.

\subsection*{Reproduction}

Reproductive investment, i.e. the biomass allocated to producing offspring (Fig. 1, C, $Z_r(t)$), depends on multiple factors, namely whether the individuals are sexually mature, the average number of offspring produced, and the weight of each individual offspring.
We thus define the biomass costs of reproduction as:
\begin{equation}
Z_{r}^{(i)}(z,t) = p_{b}^{(i)}(z) n_{o}^{(i)}(z) \mu_{o}^{(i)},
\end{equation}
where $Z_{r}^{(i)}(z,t)$ is the biomass invested in reproduction, $p_{b}^{(i)}(z)$ is the probability of reproducing, $n_{o}^{(i)}(z) \in \mathbb{R}^+$ is the average number of offspring produced, and $\mu_{o}^{(i)} \in \mathbb{R}^+$ their average body mass. 
We note that approximating reproductive investment by the total offspring biomass is a simplification. 
In reality, reproductive investment comprises additional costs, e.g. substantial energetic overheads, such as placental tissues for animals or reproductive metabolic costs, which we here assume to be small compared to the biomass of offspring produced. 

We assume that the probability of reproducing increases with body mass.
This captures indirectly the fact that individuals need to first reach sexual maturity, which happens within a range of ages and body masses that varies depending on the taxa.
This also reflects the fact that larger individuals are often more successful at reproducing, either because they are better able to secure mates in the case of animals, or because they can produce more gametes.
Therefore, we define the probability of reproducing as a sigmoid function of body mass,
\begin{equation}
p_b^{(i)}(z) = \beta_{b,0}^{(i)} \cdot \mathrm{sigmoid} \left\{\left( z - \beta_{b,1}^{(i)} \right) \beta_{b,2}^{(i)} \right\},
\end{equation}
where $p_{b}^{(i)}(z)$ is the probability of reproducing, $\beta_{b,0}^{(i)}$ is the maximum reproduction probability, $\beta_{b,1}^{(i)}$ is the body mass at reproduction onset, i.e. sexual maturity, and $\beta_{b,2}^{(i)}$ is the slope of the transition to sexual maturity. 

The number of offspring produced depends on the taxa, for instance wolves produce multiple pups per year, while elk produce one calf (\cite{Smith2020}).
We assume that successful reproducers invest a fixed fraction of their body mass at sexual maturity, and that the resulting number of offspring produced depends on their average weight.
More formally,
\begin{equation}
n_{o}^{(i)}(z) = \nu_{o}^{(i)} \frac{\beta_{b,1}^{(i)}}{\mu_{o}^{(i)}},
\end{equation}
where $\nu_{o}^{(i)} \in [0,1]$ designates the fraction of body reserves allocated to offspring production by the individuals of species $i$. 
Again, we write this quantity as a function of $z$ for the sake of generalisability of the model.

Even so, we assume that the body weight of offspring produced is never identical and falls along a distribution of possible values. 
In IPMs, this is commonly addressed by letting the phenotype of offspring be determined according to a Gaussian distribution (\cite{Lachish2020}),
\begin{equation}\begin{aligned}
& f_{r}^{(i)}(z | z') = \frac{1}{\sqrt{2 \pi {\sigma_o^{(i)}}^2}} \exp \left \{ -\frac{1}{2 {\sigma_o^{(i)}}^2} \left(z - \mu_o^{(i)} \right)^2 \right \}, \phantom{\int} \\
\end{aligned}\end{equation}
where $f_{r}^{(i)}(z | z')$ is the probability density of obtaining a offspring of phenotype $z$ given parental phenotype $z'$, and $\mu_o^{(i)}$ and $\sigma_o^{(i)} \in R+$ are the mean and standard deviation of the body mass distribution of offspring of species $i$.
In practice, to ensure that the property $\int_{S_z} f(z|z') dz = 1$ is maintained when computing the integral over a restricted range of phenotypic values, we normalise this probability density function using the sum of the finite approximation of the integral over this range.

\subsection*{Growth}

We consider here that growth is possible if there is an excess of biomass after somatic maintenance and reproductive investment (\cite{Lachish2020}).
This hence depends on the difference between the biomass gained by an individual through consumption and assimilation, $Z_a^{(i)}(z,t)$, and that which is invested into maintenance and reproduction, obtained by summing the maintenance costs and reproductive investment, $Z_l^{(i)}(z,t) = Z_r^{(i)}(z,t) + Z_m^{(i)}(z,t)$.
This difference can be added to the body mass of an individual at a time $t$ to compute its expected body mass at time $t+1$ (Fig. 1, C, $Z_g(t)$), 
\begin{equation}\begin{aligned}\label{eq:individual-biomass-balance}
& Z_{g}^{(i)}(z',t) = z' + Z_a^{(i)}(z', t) - Z_l^{(i)}(z',t). \phantom{\int} \\
\end{aligned}\end{equation}

In reality, individuals do not always achieve this expected growth due to stochasticity.
We hence assume that the phenotype after growth is determined by a normal distribution centered around the expectation of body mass change between $t$ and $t+1$ (\cite{Lachish2020}),
\begin{equation}\begin{aligned}
& f_{g}^{(i)}(z | z') = \frac{1}{\sqrt{2 \pi {\sigma_g^{(i)}}^2}} \exp \left \{ -\frac{1}{2 {\sigma_g^{(i)}}^2} \left(z - Z_g^{(i)}(z',t) \right)^2 \right \}, \phantom{\int} \\
\end{aligned}\end{equation}
where $f_{g}^{(i)}(z | z')$ is the probability density of achieving phenotype $z$ through growth given an initial phenotype $z'$, and where $\sigma_g^{(i)}$ controls the deviation from the expected body mass after growth, $Z_g^{(i)}(z',t)$.
The probability density function of growth is also normalised as explained in the previous section.

\subsection*{Biomass loss}

Another key component of this framework is the recycling of biomass through decomposition (Fig. 1, B, brown lines).
In order to determine how much biomass is recycled, we compute waste products from (i) resources not acquired following herbivory, scavenging, and predation, as well as (ii) those coming from imperfect assimilation of consumed biomass and maintenance. 

The mass of resources that is removed from a population, or from an unstructured ecosystem compartment, but not actually acquired by consumers, e.g. when a wolf (\textit{Canis lupus}) pack kills an elk but does not manage to acquire all of the available biomass, is given by
\begin{equation}\begin{aligned}
Z_{\neg c}(t) &= \sum_{i=1}^{I} \left(1 - \lambda_1^{(i)}\right) P_{d,u}^{(0,i)}(t) + \sum_{i=1}^{I} \left(1 - \lambda_2^{(i)}\right) G_{d,h}^{(0,i)}(t) + \sum_{i=1}^{I}  \left(1 - \lambda_3^{(i)} \right) \sum_{j=1}^{J} Z_{d,n}^{(j)}(t) + \\& \sum_{i=1}^{I} \left(1 - \lambda_4^{(i)}\right) \sum_{j=1}^{J} Z_{d,p}^{(j,i)}(t), \phantom{\int} \\
\end{aligned}\end{equation}
which amounts to computing the proportion of time that the consumers are spending doing any other activity than acquiring the biomass that they removed from their environment.

In addition to biomass that is not acquired, biomass can be lost because it is not successfully assimilated or because it is used for metabolic maintenance (Fig. 1, C, $Z_e(t)$).
We compute unsuccessful assimilation as the difference between the total biomass acquired and net gain in biomass.
We then add metabolic costs to compute the total loss of acquired biomass,
\begin{equation}
Z_{e}(t) = \sum_{i=1}^{I} Z_c^{(i)}(t) - \int Z_a^{(i)}(z', t) p_s^{(i)}(z') N^{(i)}(z', t) dz' + \int Z_{m}^{(i)}(z',t) p_s^{(i)}(z') N^{(i)}(z', t) dz'.
\end{equation}

\subsection*{Full ecosystem dynamics}

Now that we have computed the mass of resources removed from the environment by structured species and the associated waste-products of these activities, we can write the full ecosystem dynamics. 
In addition to the biomass distribution of structured plant and animal species, we assume that the state of the full ecosystem also depends on the biomass of organic matter, i.e. partially degraded organic compounds, the biomass of decomposer communities, i.e. organisms that consume this organic matter such as invertebrates and bacteria, thereby contributing to the release of nutrients (e.g. nitrogen and phosphate), and the biomass of unstructured plant communities, that we assume here to be solely grass species (Fig. 1, B, bottom brown lines). 
We model the year-to-year change in the total mass of these ecosystem compartments and their inter-dependencies using a system of difference equations derived from simplifying equations in \textcite{Chertov1997}:
\begin{equation}\left.\begin{aligned}\label{eq:de-unstructured}
D(t+1) &= D(t) + \phi_{BD} B(t) + \phi_{GD} G(t) - h(\phi_{DB} D(t), \rho_B B(t)) + Z_{\neg c}(t) + Z_e(t), \phantom{\int} \\
B(t+1) &= B(t) + (1 - \phi_{DP}) \cdot h(\phi_{DB} D(t), \rho_B B(t)) - \phi_{BD} B(t), \phantom{\int} \\
P(t+1) &= P(t) + \phi_{DP} \cdot h(\phi_{DB} D(t), \rho_B B(t)) - h(\phi_{PG} P(t), \rho_G G(t)) - P_{d,u}(t), \phantom{\int} \\
G(t+1) &= G(t) + \gamma_3 \cdot h(\phi_{PG} P(t), \rho_G G(t)) - \phi_{GD} G(t) - G_{d,h}(t), \phantom{\int} \\
\end{aligned}\right\}\end{equation}
where $D(t) \in \mathbb{R}^{+}$ is the biomass of organic matter, $B(t)\in \mathbb{R}^{+}$ is the biomass of the community of decomposers, $P(t)\in \mathbb{R}^{+}$ is the mass of nutrients (e.g. phosphates or nitrogen), $G(t)\in \mathbb{R}^{+}$ is the biomass of grass species.
Nutrient recycling is achieved by introducing biomass input terms in the dynamics of the organic matter and removal terms in the dynamics of nutrients and grass.
Therefore, the dynamics of the organic matter, $D(t)$, are increased by waste products of consumers, $Z_{\neg c}(t)$ and $Z_e(t)$, and the dynamics of nutrients and grass are reduced by the mass of resources acquired by consumers, $P_{d,u}(t) = \sum_i P_{d,u}^{(0,i)}(t)$ and $G_{d,h}(t)= \sum_i G_{d,h}^{(0,i)}(t)$, respectively.

The parameters $\phi \in [0, 1]$ denote proportions, to prevent exceeding the maximum available pool of resources it relates to, while the parameters $\rho \in \mathbb{R}^{+}$ denote rates, which can be greater than one.
The function $h(x,y) = x y / (x + y)$ is the half harmonic function, which is equivalent to the Michaelis-Menten equation for fixed values of x.
This function ensures that the mass consumed cannot exceed the mass available for consumption, and reduces the per-capita consumption with increasing density of consumers.

The parameters $\phi_{BD}$ and $\phi_{GD}$ correspond to the proportions of decomposer biomass and grass biomass that are converted into organic matter following death.
The parameters $\phi_{DB}$ and $\rho_{B}$ correspond to the proportion of the biomass of organic matter available to decomposers and the rate at which it is consumed, respectively.
This construction assumes that the organic matter is consumed by decomposers, but that only a fraction $1 - \phi_{DP}$ is effectively assimilated due to imperfect uptake (e.g. external digestion by bacteria), releasing a fraction $\phi_{DP}$ as nutrient byproducts. 

This representation of nutrient recycling through decomposition of organic matter, or detritus (\cite{Schmitz2015}), is a simplification of a more complex set of processes. 
Our implementation only considers microbial communities, which control the rate of detritus decomposition via exoenzyme production (\cite{Schmitz2015}), and captures a part of the nutrients contained in the soil for their own metabolism, similar to the model of \textcite{Chertov1997}. 
It does not account for feedbacks within soil communities that can be caused by top down effects of microbivores on bacterial communities, or by fluctuations in nutrient content of organic matter which may affect bacterial growth and response to consumption (\cite{Schmitz2015}).
This representation of the nutrient cycle nonetheless captures the feedback between consumers and plants mediated by decomposers.

Finally, $\phi_{PG}$ and $\rho_{G}$ correspond to the nutrient mass available to grass and the nutrient uptake rate.
The term $\gamma_3 \in R^{+}$ is a scaling factor to reflect the fact that nutrient uptake only accounts for a fraction of the grass biomass produced, as it also depends on light and carbon intake through photosynthesis.
Overall, this system achieves fluxes of biomass from organic matter to decomposers, a fraction of which is diverted into into nutrients, then into grass, and finally back to organic matter through mortality of decomposers and grass.

\subsection*{Computational implementation}

We implement the model in PyTorch, which is a deep learning library for python.
This choice is motivated by the gain in computation speed achieved by PyTorch over R.
We estimated that PyTorch was 8.3 times faster than R for simulating IPMs in a preliminary analysis.
This gain in speed is attributable to the optimisation of matrix operations in PyTorch.
The general Eco-IPM is defined as a class with parameters coded as attributes and loaded from a YAML file.
This parameter file contains the name and value of all parameters.
Structured species can be added or removed from the model by adding or removing a species columns to the parameter file. 
This makes our framework highly flexible and easily applicable to any ecosystem with sufficient information regarding the survival, reproduction, and growth of the key species present.

To run the Eco-IPM, one needs to provide parameter values as well as initial conditions for all state variables, i.e. $D(t=0)$, $B(t=0)$, $P(t=0)$, $G(t=0)$, and $N^{(i)}(z,t=0)~\forall~i$.
We explain in the case study below how to select parameters and initial conditions for the northern Yellowstone ecosystem.
Simulating the model then proceeds by computing the year-to-year change in structured species (Eq. \ref{eq:ipm}) first, and then in the unstructured compartments (Eq. \ref{eq:de-unstructured}), though the order does not matter.
At the end of the simulation, the model returns a time series of all the variables, which are either time series of total mass for unstructured compartments, or time series of biomass distributions for the structured species.

\section*{\uppercase{Case study: dynamics of northern Yellowstone}}

In this section we showcase how the Eco-IPM framework can be applied to a real system. 
We use the case study of vegetation and large mammals in northern Yellowstone.
We explain how to define the model for northern Yellowstone, how to determine initial parameter values, and finally how to estimate those parameters from time series data of animal counts.

\subsection*{Aims}

The general aim of this case study is to demonstrate that Eco-IPMs can be parameterised a priori from the literature and calibrated with real data, such that they produce biologically plausible dynamics for the ecosystem that they are emulating.
Plausibility will be evidenced by showing (i) long-term coexistence of key species that are present in the real ecosystem, (ii) similarities between organismal properties predicted by the model, such as average body mass of adults, and their expected values known from the literature, and (iii) dynamical responses of the model to perturbations that are consistent with the real system, by taking the example of the predator extirpation and recovery that occurred in Yellowstone.

\subsection*{System}

So far, we have designed a general ecosystem model with structured and unstructured species, i.e. depending on whether we model their mass distributions or not. 
Here we show how this model is adapted to portray the dynamics of a real system.
We focus on northern Yellowstone, an area of 995 km$^2$ located along the northern boundary of Yellowstone National Park that has been extensively monitored over the past century and has produced count data for key species.
Intensive monitoring of populations in the park yielded count estimates for elk (\textit{Cervus canadensis}), bison (\textit{Bison bison}), wolves (\textit{Canis lupus}), and cougars (\textit{Puma concolor}), from the point of reintroduction of wolves in the mid 1990s to today.
Vegetation, and particularly willow (\textit{Salix spp.}) and aspen (\textit{Populus tremuloides}), also received substantial attention, although they were mostly studied in detail in very localised plots.
This system has been severely impacted by anthropogenic stressors.
By the 1930s, wolves had been extirpated from Yellowstone and cougars had been greatly reduced and probably also extirpated (\cite{Skinner1927, Weaver1978}).
From 1935 to 1968 hunters in the State of Montana harvested about 43,700 elk, with most harvest concentrated along the northern boundary as elk migrated outside the park. 
In response to concerns about overgrazing and damage to deciduous woody plants, including aspen and willow, rangers removed an additional 26,200 elk from the park through shooting or live capture and translocation (\cite{Houston1982}).
From 1969, park managers ended elk removals inside the park and allowed the elk population to respond to forage availability, harvests outside the park, predation (bears, coyotes, cougars starting in the 80s), and weather. 
Without direct population control inside the park, the herd increased substantially, peaking at approximately 19,000 elk in 1994 (\cite{Smith2020}).
The reintroduction of wolves to Yellowstone National Park in 1995--1997 has coincided with decrease in elk numbers to only a few thousands, and aspen and willow are showing signs of regeneration (\cite{Hobbs2024}).
Yet, quantifying precisely the importance of the role played by wolves and cougars in the Yellowstone trophic cascade is an ongoing challenge as confounding factors may have also indirectly contributed to vegetation regeneration by promoting elk decline, such as hunting outside of the park, predation by other species than wolves and cougars, and climatic drivers that influence winter severity, droughts, and wildfires.
Nonetheless, this system offers a unique opportunity to monitor impacts of reintroduction of large carnivores on large herbivores and vegetation. 

\subsection*{Model definition}

Our framework is designed so that organic matter, nutrients, decomposers, and unstructured plants, such as grass, are always present in the ecosystem.
Applying our model to a specific ecosystem requires adding a selection of the key structured plant and animal species that are also present in the ecosystem.
Key species of interest in northern Yellowstone are willow and aspen for woody deciduous plants, elk and American bison for large herbivores, and wolves and mountain lions for large carnivores, all modelled as structured variables.
The main equation (Eq. \ref{eq:ipm}), defined in the model description section, applies to any structured plant or animal species. 
Differences between primary producers, herbivores, and predators, are hence achieved solely through different choices of parameters.
Therefore, adding a species to the ecosystem model amounts to defining a set of parameters and initial conditions for that species.
In practice, this is done by adding columns for new structured species in the parameter file.

\subsection*{Parameterisation \textit{a priori}}

Parameterisation of the dynamics of structured and unstructured variables can be achieved by acquiring initial values for all of these parameters from the literature or from experts.
For the four well-studied large mammal species, we relied on dedicated textbooks (\cite{White2015, Smith2020, Ruth2019, White2024}), which yielded most of the necessary estimates.
Those were complemented by targeted literature searches on Google Scholar for specific parameter values that were not found in the textbooks.
We followed this procedure for willow and aspen, and unstructured compartments.
We identified a total of 77 academic sources from which we derived estimates for the 144 parameters of the model for the four unstructured compartments and six structured species.
We provide a summary of our initial parameter values in Table S2--4 (see supplementary section S3), and multiple supplementary section describing in detail how we derived them from the literature (see supplementary sections S5--12).

Overall, 138/144 of these initial estimates were satisfactory in the sense that we did not need to modify them to obtain suitable ecosystem dynamics, i.e. model simulations where the ecosystem reaches an equilibrium wherein all species coexist.
Our first estimates of metabolic costs derived from the literature were too high compared to resource consumption in the equation that determines the balance between gains and losses (Eq. \ref{eq:individual-biomass-balance}), possibly due to the numerous steps that can lead to biomass loss from capture in the environment to successful assimilation.
We therefore scaled down by hand the metabolic cost coefficients of all six species uniformly to achieve coexistence of the six structured species.
Scaling was done by repeatedly halving the metabolic coefficients until the model produced simulation with non-zero population sizes.
This yielded a scaling factor of $0.125$.
This is the only modification that was necessary to obtain a suitable initial model for this simplified northern Yellowstone ecosystem.

\subsection*{Initial conditions}

In addition to setting the initial value of the parameters, simulating the model also requires the specification of initial conditions for unstructured variables, i.e. $D(t=0)$, $P(t=0)$, $B(t=0)$, $G(t=0)$ and structured variables, $N^{(i)}(t=0,~z),~i=1,...,6$. 
The indices $i=1,...,6$ denote in order willow, aspen, elk, bison, wolves, and cougars.
A simple way to determine the initial value of variables in the absence of data, is to initiate the model using arbitrary initial values, such as Gaussian distributions for biomass distributions of the species, 
\begin{equation}
N^{(i)}(t=0,~z) = N^{(i)}(t=0) \times \mathrm{Normal}\left(\mu_0^{(i)}, \sigma_0^{(i)}\right), 
\end{equation}
where $N^{(i)}(t=0)$, $\mu_0^{(i)}$, and $\sigma_0^{(i)}$ are arbitrary values defined by the user, and then simulate the model until it reaches an equilibrium.
The equilibrium biomass distributions of structured species and point values of unstructured variables can then be used as a starting point for further simulation or perturbation analyses.

Once initial parameter values and initial conditions have been determined, the model can be simulated as a regular difference equation system, that is, by sequentially computing the change in each variable of the system between the current and the next time step, before updating their value and setting those as the starting point for the next time step.
In practice, we start by computing year-to-year change in structured populations, $N(t+1,~z)^{(i)}$, and then unstructured populations, $D(t+1)$, $P(t+1)$, $B(t+1)$, and $G(t+1)$, but the order does not affect computations.

To fit the model to the time series of animal counts in northern Yellowstone, we select initial conditions that are as close as possible to the state of the system in 1995, when the wolves were reintroduced.
Counts of animals were known for that year except for elk and cougars, for which the nearest counts available were for 1994 and 1999, respectively.
We thus set the count of animals to the count for that year, or closest available time point.
Aside from these counts, we found little information regarding the initial state of other ecosystem components, i.e. individual counts and biomass distributions for willow and aspen were not available, and no information exists regarding the total mass of organic matter and nutrients and total biomass of decomposers and grass in northern Yellowstone.
This is commonly addressed by priming the model with initial states inferred from starting simulations prior to the fitting time period (\cite{Elderd2016, White2016}). 
Therefore, we prepared the model by emulating the human interventions that happened in northern Yellowstone prior to wolf reintroduction.
First, we simulated the model for 200 years to obtain equilibrium states for all variables.
Second, we removed predators by setting their population size to zero and simulated the model for 25 years, which is approximately the time period during which no culling took place in northern Yellowstone (i.e. 1970--1995). 
Finally, we set initial conditions to the values thereby obtained, which included the total mass in each unstructured compartment, the biomass distributions and densities of aspen and willow, and the biomass distributions of animals, as their densities could already be estimated from the time series data. 

\subsection*{Model calibration}

We use Bayesian inference to fit the model to the time series of animal counts in northern Yellowstone.
This is because the literature estimates may not necessarily lead to dynamics that resemble that of the real system.
In practice, this amounts to correcting the value of the literature estimates such that the model predictions better agree with the time series data collected for this system.
To do this we define a simple Bayesian model (\cite{Hobbs2015}) and perform Bayesian regularisation (\cite{Cawley2007}).
We designate the expected value of the $k^{th}$ parameter obtained from the literature as $\omega_{k}$.
We introduce a scaling parameter $\alpha_k \in \mathbb{R}$ which we use to scale up or down the corresponding literature estimate, which results in a corrected estimate for that parameter,
\begin{equation}\left.\begin{aligned}
& \theta_k := \alpha_k~\omega_k, \\
& \alpha_k ~\overset{\text{iid}}{\sim}~ \text{Normal}(1, \sigma_{\alpha}),
\end{aligned}\right\}\end{equation}
where $\theta_k$ is the corrected estimate and $\sigma_{\alpha}$ is a hyperparameter that controls the general level of uncertainty around the literature estimates.
For instance, $\sigma_{\alpha} = 0.1$ encourages on average a deviation of $10\%$ from the literature estimates.
More conservative inference can be performed by using smaller values of this coefficient, which enforces values of the scaling parameters close to one, and therefore close to the literature estimate.
Conversely, larger values of the regularisation coefficient will enable larger deviations of the corrected estimates from the literature estimates.
We note that this can lead to potential flips in the sign of the parameters, which may not be biologically meaningful for some parameters. 
In practice, keeping the value of $\sigma_\alpha$ small ensures that these occurrences are highly unlikely.
For parameters with support on $[0,1]$, such as the asymptotic probability of survival, we scale the log-odds rather than the actual value of the parameter, 
\[
\theta_k := \text{sigmoid}\left\{\alpha_k \cdot \log \left\{ \frac{\omega_k}{1-\omega_k} \right\}\right\},~\text{if}~\omega_k~\in~[0,1].
\]
This ensures that those parameters remain on the interval $[0,1]$ after scaling.

We define a Gaussian likelihood for the observed population density estimates given the model predictions,
\begin{equation}
N^{(i)}_t ~\overset{\text{iid}}{\sim}~ \text{Normal} \left( N^{(i)} ( t; \theta^{(i)} ), \sigma^{(i)}_{N} \right),
\end{equation}
where $N^{(i)}_t$ is the observed  density of the $i^{th}$ species at time $t$, $N^{(i)}(t; \theta^{(i)})$ is the corresponding model prediction given the corrected parameter vector $\theta^{(i)}$, and $\sigma_N^{(i)}$ is the observation error.
We note that we assume here that population densities are normally distributed, though they are variables with positive support. 
This is a convenience assumption which enables mean squared error optimisation after log-transformation of the posterior. 
In practice, it is unlikely to pose a problem because observations of population counts in the time series were orders of magnitude larger than zero.

The full posterior density distribution can thus be written by combining the likelihood and prior density of parameters for each species for which time series are available,
\begin{equation}\begin{aligned}
p \left(\theta^{(1)}, \ldots, \theta^{(I)}, \sigma_N^{(1)}, \ldots, \sigma_N^{(I)} ~|~ N^{(1)}_{t}, \ldots, N^{(I)}_{T} \right) \propto \prod_{i=1}^{I} \prod_{t=0}^{T} p\left(N^{(i)}_t ~|~ N^{(i)} ( t; \theta^{(i)} ), \sigma^{(i)}_{N} \right) p\left(\theta^{(i)}\right),
\end{aligned}\end{equation}
where $p(.)$ denotes probability density functions, $I$ the total number of species, and $T$ the number of time steps. 
For the sake of simplicity, we write the posterior distribution as a function of the corrected parameters, $\theta$. 
However, in effect the objects of the inference are the scaling parameters $\alpha$.
Those are not represented, due to a deterministic relationship between $\theta$ and $\alpha$ which implies that $p(\theta) = p(\alpha)$.
Therefore, calibrating the model, i.e. fitting it to the time series, consists in finding the values $\alpha_k$ that maximise the posterior density.

We did not have time series of willow and aspen population counts, as most data available comes from measuring a subset of individuals in selected stands and focuses on new growth, thus not accounting for older trees. 
Therefore, the posterior is optimised only considering time series of elk, bison, wolf, and cougar counts, i.e. $i = 3,...,6$.
The time series span 30 years, from 1995 to 2025.
Missing counts were computed as NAs and discarded from the evaluation of the posterior density.
To avoid differences in the weight of a time series in the likelihood computation due to different numbers of observations, we weighted the likelihood of each species using the inverse of the number of observations.
Given that we rely on Gaussian likelihoods, this amounts to computing the mean squared error rather than sum of squares following log transformation. 
Each time series was standardised by computing the standard score, $z = (x - \mu)/\sigma$, where $\mu$ and $\sigma$ are the mean and standard deviation of the variable across the entire time series.
Counts predicted by the model were standardised following the same procedure, using the mean and standard deviation of the time series.
Standardisation was only used for evaluating the posterior, variables were brought back to their natural scale for all simulations and downstream analyses.

We maximise the posterior distribution with respect to $\alpha$ using a gradient-free optimisation algorithm based on single-chain differential-evolution Monte Carlo (DEMC) (Bonnaffe 2022), derived from multi-chain DEMC (\cite{TerBraak2006}).
This algorithm collects iteratively parameter values associated with an increase in posterior density, thereby forming a chain. 
The next element of the chain is proposed by computing the difference between two previous states selected at random.
Noise is applied to improve the diversity of solutions proposed.
The proposed element is accepted if it leads to an increase in the posterior density.
Using larger values for the hyperparameter controlling the amplitude of the noise usually leads to faster identification of the appropriate scale of variation of the parameters.
The problem is that the noise level usually needs to be scaled down as the dimensions of the parameter space increase. 
This limits the applicability of this approach for the present models which have many more parameters than the dynamical models used in the development of the methodology.
To circumvent this problem we introduce a sparse noise parameter, which enables larger perturbations in a subset of the variables.
The algorithm is described more formally in Table 2.
This algorithm was a necessary extension of the original framework to deal with the relatively higher dimensionality of the parameter space of the models that we consider in the present work.
To account for the existence of multiple solutions, and identify the global optimum, we repeat the fitting procedure multiple times (N = 30) and retain the parameter vector associated with the overall highest posterior density.
To assess robustness of our results to alternative solutions, we ensured that parameter scalings, model dynamics, and main findings were consistent across the three best models and a model trained on a fraction of the time series (Fig. S2--6).

\begin{table}
\centering
\caption{\textbf{Single-chain differential-evolution optimisation algorithm with sparse noise.}}
\setstretch{1.0}
\begin{tabular}{r p{14cm}}
\toprule
1  & Initialise $k = 0$ \\
2  & \quad Choose $\theta^{(0)}$ such that $\pi(\theta^{(0)}) > 0$ \\
3  & \quad Set $\gamma = 2.38/\sqrt{2d}$, $\delta = 0.001$, $\zeta = 0.1$, $\varphi = 0.1$, $\lambda = 25$, $k_{max} = 100$ \\
4  & \textbf{for} $k = 1, \ldots, k_{\max}$ \textbf{do} \\
5  & \quad Sample $u, v \sim \text{Uniform}\{k-\lambda,\dots,k-1\}, \ u \neq v$ \\
6  & \quad Draw $\epsilon \sim \text{Uniform}(-\delta,\delta)$, $\eta \sim \text{Uniform}(-\zeta,\zeta)$ \\
7  & \quad Let $\beta = (\beta_1,\dots,\beta_d),\ \beta_i \overset{\text{iid}}{\sim} \text{Bernoulli}(\varphi)$ \\
8  & \quad Compute $\theta_p = \theta^{(k)} + \gamma(\theta^{(u)} - \theta^{(v)}) + \epsilon + \eta\beta$ \\
9  & \quad \textbf{if} $\pi(\theta_p) > \pi(\theta^{(k)})$ \textbf{then} set $\theta^{(k+1)} = \theta_p$ \\
10 & \quad \textbf{else} set $\theta^{(k+1)} = \theta^{(k)}$ \\
11 & \textbf{return} $\Theta = (\theta^{(0)},\dots,\theta^{(k_{\max})})$ \\
\bottomrule
\end{tabular}
\end{table}

We also introduce an early stop condition following which a given optimisation run stops if the parameter vector remains unchanged after a set number of iterations.
Each training epoch consists of 100 steps of the optimiser. 
Early stopping comes in effect if the log-posterior density does not increase by more than 0.0001 across three consecutive epochs. 
The optimisation will thus continue unless the parameters have remained unchanged for 3 epochs, that is 300 iterations, by default.
We determine suitable values for the hyperparameters controlling the noise level, $\epsilon$, $\zeta$, and $\varphi$, by trial and error on shorter optimisation runs with 10 epochs.
If noise levels are set too high, they will cause changes to the parameter vector that are too large, and therefore will lead to implausible model simulations, thus resulting in an early termination of the optimisation due to the chain getting stuck. 
If noise levels are too low the optimisation will progress slowly. 
The acceptance rate, i.e. the number of proposed parameter vectors accepted, is a good indicator of the quality of the optimisation and should remain around 30\% to guarantee efficient optimisation.
We also proceed by trial and error to identify the appropriate regularisation parameter, $\sigma_\alpha$.
We tested $\sigma_\alpha = 0.1$ and run 3 pilot optimisation runs.
We repeated this for $\sigma_\alpha = 1.0$ and $\sigma_\alpha = 2.0$.
Setting $\sigma_\alpha = 0.1$ led to consistent underfitting, whereby the model would capture the average population counts but not the dynamics.
Setting $\sigma_\alpha = 2.0$ yielded overfitting, which caused the optimiser to get stuck in local maxima. 
The value of $\sigma_\alpha = 1.0$ produced smooth optimisation runs, which we used for the full optimisation sequence.

To provide further evidence that the fitting algorithm works as intended, we also fitted the model to synthetic data.
The synthetic dataset was obtained by perturbing the initial parameter scalings with Gaussian noise using a standard deviation of $\sigma_{noise} = 0.1$, which corresponds to a $10\%$ deviation from the literature estimates on average across all parameters. 
We then simulated the model for 30 time steps, using the same initial conditions as for fitting the model to the real time series.
Finally, we ran 20 optimisation runs, which was sufficient to identify multiple near-perfect fits, and selected the parameter set that maximised the posterior density.
The fit of the best model to the simulated data is reported in the first section of the results.

\subsection*{Model validation}

Even if the Bayesian regularisation approach that we employ here reduces the chances of overfitting, it does not completely eleminate this possibility.
Overfitting can generally be evidenced by the model not being able to generalise to new data points.
In our case, this could manifest as the model not being able to predict correctly the future state of the ecosystem.
So, to further assess the robustness of our results and analyses to overfitting, we also perform out-of-sample testing, whereby we train our model on only a fraction of the time series. 
The first 20 years were used for training 100 models. 
The years 20 to 25 were used to identify the model with the highest generalisation capacity. 
Finally, the last 5 years of the time series were used as an ultimate test of the predictive performance of the best model.
Good performance on the test set would be evidence that the model has learned a general dynamical process.
Furthermore, finding congruent parameter values and dynamical patterns between the model trained on the full time series and the model trained only on a fraction of the time series would be a good indication that those results are not the product of overfitting.
Therefore, we compare all the results derived from the model trained on the full data to those derived from the model which only has learned from a fraction of the time series (Fig. S2--6).
We still use the model trained on the full time series for inferences as it has learned from more data and may thus yield more accurate representations of the underlying processes.

\subsection*{Model analyses}

Though the calibration step is not required in order to analyse the model, as the model can be simulated to observe transient dynamics and equilibrium points, and investigate how these are affected by a change in the parameters, the calibrated model has the advantage of behaving similarly to the real system, and can thus be used as a digital surrogate of its real counterpart.
The full range of analyses that can be done with these models extends far beyond those considered in this work and are presented in details in a dedicated textbook (\cite{Ellner2016}).
In this study, we analyse the model in three ways, to verify that Eco-IPMs satisfy the three expectations that we formulated in the aims section. 

First, we observe long term dynamics that emerge from simulating the model for an extended period of time.
The aim of this analysis is to determine whether there exists an equilibrium state in our model of northern Yellowstone where all species coexist, and what ecosystem state this corresponds to in terms of species biomass distributions and densities. 

Second, we assess whether some of the biological properties predicted by the model, but not explicitly enforced by specific parameters, are plausible. 
We compute the average body mass of adults of each animal species and compare it to the values expected from the literature.
Knowing the body mass at maturity for each species, we can compute the expectation of the body mass of individuals larger than this threshold to approximate the mean body mass of adults:
\begin{equation}
\bar{z}^{(i)}_{z\geq\beta^{(i)}_{b,1}}(t) = \frac{1}{N^{(i)}(t)}\int_{\beta^{(i)}_{b,1}}^{\infty} z N^{(i)}(z,t)dz.
\end{equation}

Third, we determine whether the response of the model system to an external perturbation is consistent with that of the real system. 
In particular, we seek here to see if the model responds in a plausible way to the removal of predators, though it was only trained to respond to predator reintroduction.
To demonstrate this, we perform an artificial predator extirpation and reintroduction experiment to determine the impact of predators on other ecosystem states as well as disruption and recovery times.
We initialise the experiment by setting the variables at a stable equilibrium point, identified by performing a 200-year-long simulation of the fitted model. 
The experiment can be divided into three periods of equal lengths.
In the first period, we simulate the system for 200 years from its initial equilibrium state to demonstrate stability in initial conditions.
In the second period, we remove the two predators, wolves and cougars, by setting their population size to zero, then simulate the system for a further 200 years.
In the third period, we reintroduce 100 individuals in the wolf and cougar populations, considering a normal distribution of body mass centered around 40 kg with a standard deviation of 1 kg, similar to that of wolves that were reintroduced to Yellowstone in 1995 (\cite{Smith2020}).
The system is then simulated for a further 200 years.
In total, the experiment spans 600 years.
The second and third periods show the response of the system to a human-induced disruption and its capacity to return to its original state.
We quantify transient dynamics for the disruption and recovery periods, namely the time that it takes each species to attain an equilibrium point whereby time-averaged population sizes are constant through time.
We also compare the state that each population is in at the end of the third period to the initial period to determine whether the recovery was complete or not.
We performed the exact same artificial experiment with disruption and recovery periods set to 1500 years to capture potential long term effects, such as local species extinctions.
The results of the long term extirpation experiment are presented in the supplementary material (see supplementary section S2).

\paragraph{Elasticities} As with IPMs in general, it is possible to calculate elasticities of a component of the ecosystem on another, namely the change of a target quantity, expressed as a proportion of its current value, with respect to a proportional change in another variable or parameter in the model. 
Those are computed as follows:
\begin{equation}
\epsilon_{x \rightarrow y}(x, y) = \frac{\partial y}{\partial x} \frac{x}{y},
\end{equation}
where $y$ is the target quantity, and $x$ is the variable with respect to which the elasticity is calculated.
We compute elasticities in our predator extirpation and recovery experiment to determine how the strength of resource limitations on plant and herbivore growth varies with presence or absence of their respective predators.
More specifically, we calculate $\epsilon_{Z_c(t) \rightarrow Z(t+1)}$ at each time step in the extirpation-recovery experiment, which is the proportional change in annual biomass growth of a population, $Z(t+1)$, with respect to a change in total acquired biomass, $Z_c(t)$.

\section*{\uppercase{Results}}

\paragraph{Model fit to the simulated time series} 
Figure 2 shows the results of fitting the model to the simulated time series. 
Panel A reveals that the log likelihood is optimised rapidly, contrary to the log priors, which means that regularising the parameters, such that they are closer to the literature estimates, constitutes the longest part of the training.
Panel B shows that the initial parameter values, used to generate the artificial time series, are correctly retrieved within 120 epochs.
Panel C--F show that the predicted counts match the artificial observations.

\begin{figure}
  \centering  
  \includegraphics[width=\textwidth, page=2, trim={0 1cm 0 1cm}, clip]{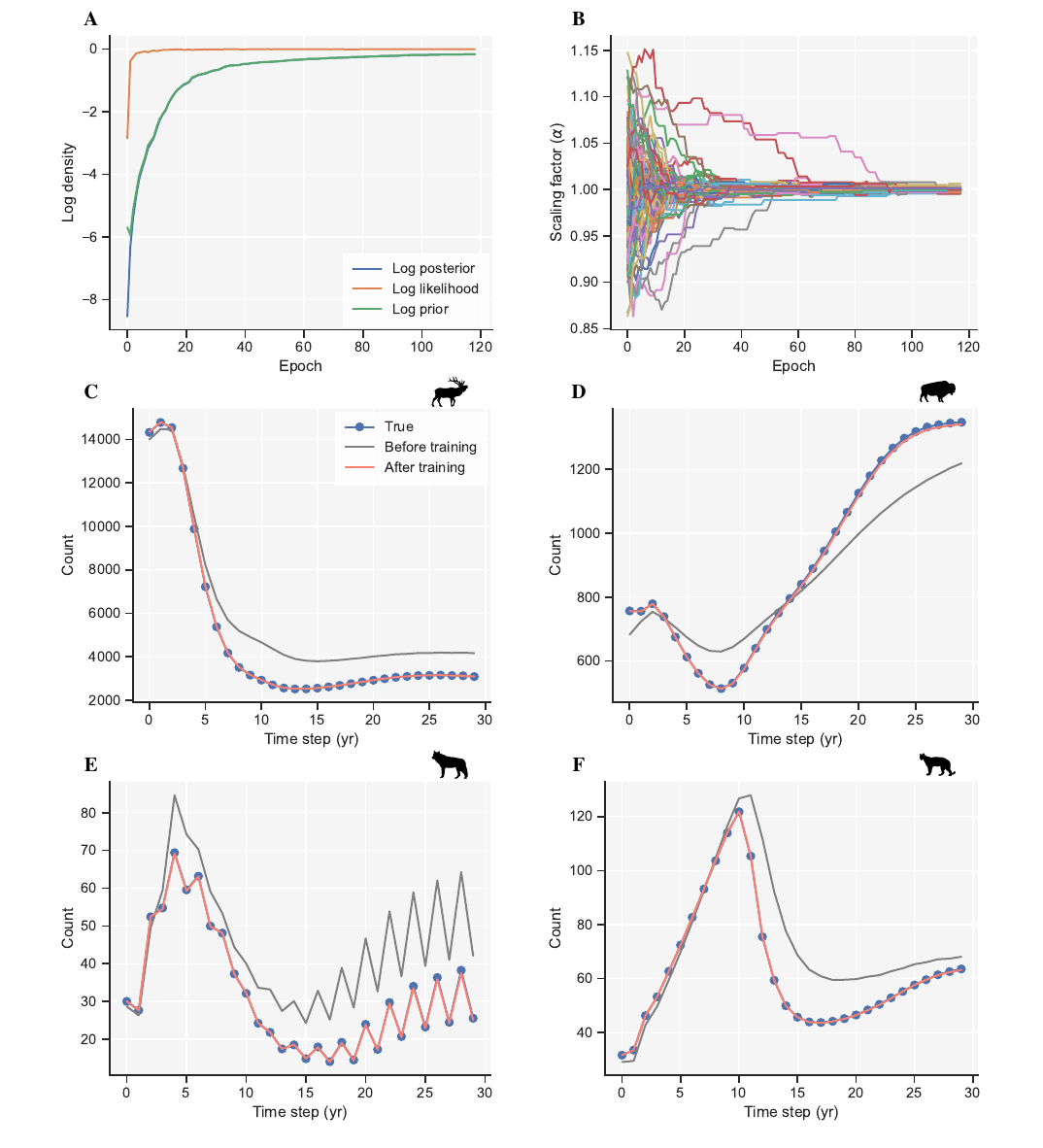}  
  \caption{
    \textbf{Yellowstone ecosystem model fit to the simulated time series data.}
    Panel A shows the change in log posterior density, likelihood, and prior density with the training epoch.
    Each epoch consists of 100 steps of sparse noise differential-evolution Monte-Carlo.
    Panel B displays the trace of each element of the parameter vector, i.e. the scaling factors that upscale or downscale the prior estimates.
    Panel C--F show the predictions of the model before and after training, compared to the artificial data simulated by the model using parameter values \textit{a priori}, i.e. derived from the literature.
    The fit before training uses perturbed parameter values, i.e. with an added Gaussian noise of amplitude $\sigma_{noise} = 0.1$.
    The fit after training uses the parameter vector that maximises the log posterior density, selected from a sample of $N = 20$ optimisation runs.
    The initial conditions are set to match the observed count, or the closest available estimate (in time).
    Figure and icons created by Willem Bonnaff\'e.
  }
  \label{fig:model-fit-simulated}
\end{figure}

\paragraph{Model fit to the real time series}
Figure 3 presents the results of the fit of the model to the real time series.
The fit of the model before training, namely with literature-derived estimates, already reproduces the main trends observed in the time series, in spite of a tendency to underestimate elk, bison, and wolf counts, and overestimate cougars (Fig. 3, A--D, grey line).
After training, the model captures the overall numbers and the demographic trends for all species (Fig. 3, A--D, red line).
The main discrepancy between observed counts and model predictions consists of a slight underestimation of elk and bison counts at the end of the time series (Fig. 3, A--B, red line, $> 20$ years), and a slight overestimation of bison counts at the start ($< 10$ years).
We also note that large fluctuations in wolf counts are not captured by the model (Fig. 3, C).

\begin{figure}
  \centering    
  \includegraphics[width=\textwidth, page=3, trim=0 2cm 0 2cm, clip]{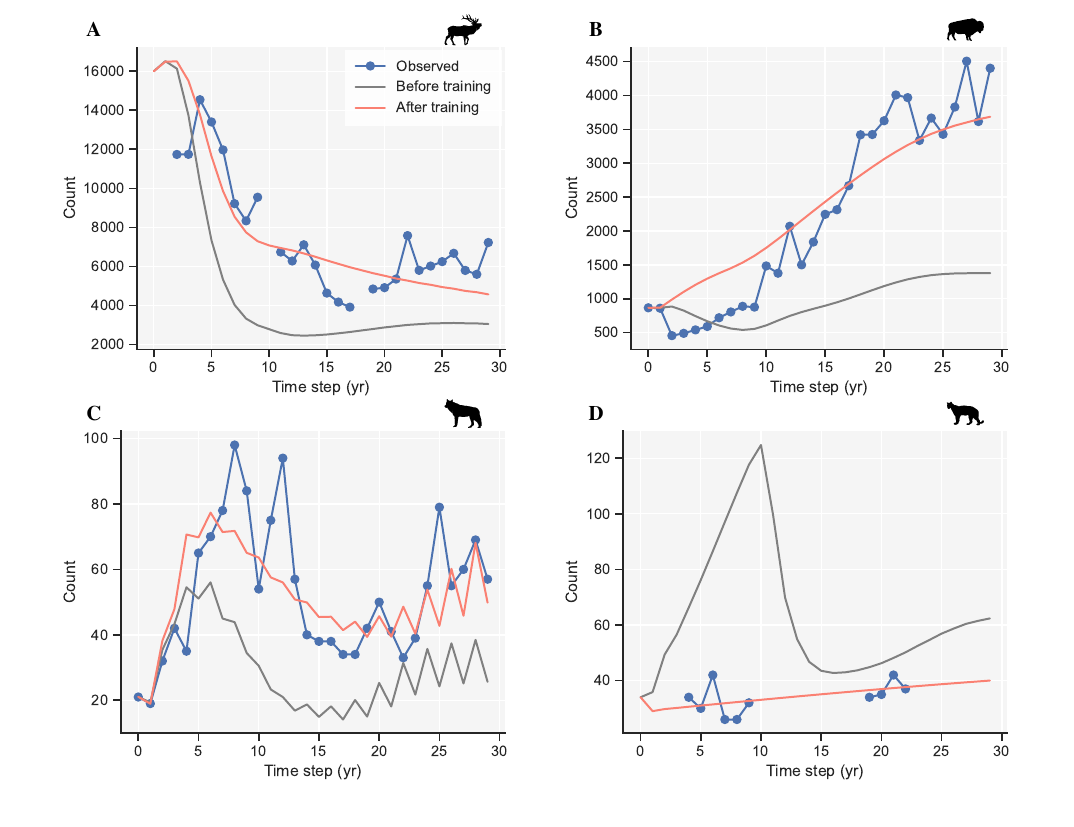}
  \caption{
    \textbf{Yellowstone ecosystem model fit to the real time series data.}
    Panel A--D show the predictions of the best model (out of 30 models) before and after training, compared to the count data collected in northern Yellowstone, for elk, bison, wolves and cougars, respectively.
    The fit before training uses the parameter values \textit{a priori}, i.e. derived from the literature.
    The fit after training uses the parameter vector that maximises the log posterior density, selected from a sample of $N = 30$ optimisation runs.
    The initial conditions are set to match the observed count, or the closest available estimate (in time).
    Figure and icons created by Willem Bonnaff\'e.
  }
  \label{fig:model-fit-full}
\end{figure}

\paragraph{Results of the optimisation}
Figure 4 presents the results of the optimisation of the log posterior density of the parameters given the time series data.
Most optimisation runs initially display a sharp increase in likelihood and a coincidental decline in prior density, but quickly revert to a joint increase (Fig. 4, A, orange and green line).
This is mirrored by the initial spread of the scaling factors, followed by a progressive regularisation of most estimates towards one (Fig. 4, B).
At the end of optimisation, most parameter values converge back to one, with only $N = $13 parameters up- or downscaled by more than 2.5\% of their initial value (i.e. estimated from the literature) (Fig. 4, C).
Parameters that are downscaled are the body mass at sexual maturity, $\beta_{b,1}^{(6)}$, and reproductive investment of cougars, $\nu_o^{(6)}$, as well as the body mass of wolf pups, $\mu_o^{(5)}$, the metabolic cost scaling coefficient of elk and bison, $\delta^{(3)}$ and $\delta^{(4)}$ (Fig. 4, D, blue dots, indices in the figure and script start at 0).
Parameters that are upscaled are the biomass assimilation efficiency of elk and bison, $\alpha_1^{(3)}$ and $\alpha_1^{(4)}$, the mass at reproduction onset in wolves, $\beta_{b,1}^{(5)}$, reproductive investment, $\nu_o^{(5)}$, and slope of intra-specific mortality, $\beta_{d,p,2}^{(5,5)}$, the proportion of nutrients available to grass and its nutrient efficiency, $\phi_{PG}$ and $\gamma_3$, and the mass of cougar offspring, $\mu_o^{(6)}$ (Fig. 4, D, red dots).
More details are provided in the supplementary material (see supplementary section S3, Table S1).

\begin{figure}
  \centering  
  \includegraphics[width=\textwidth, page=4, trim={0 2cm 0 2cm}, clip]{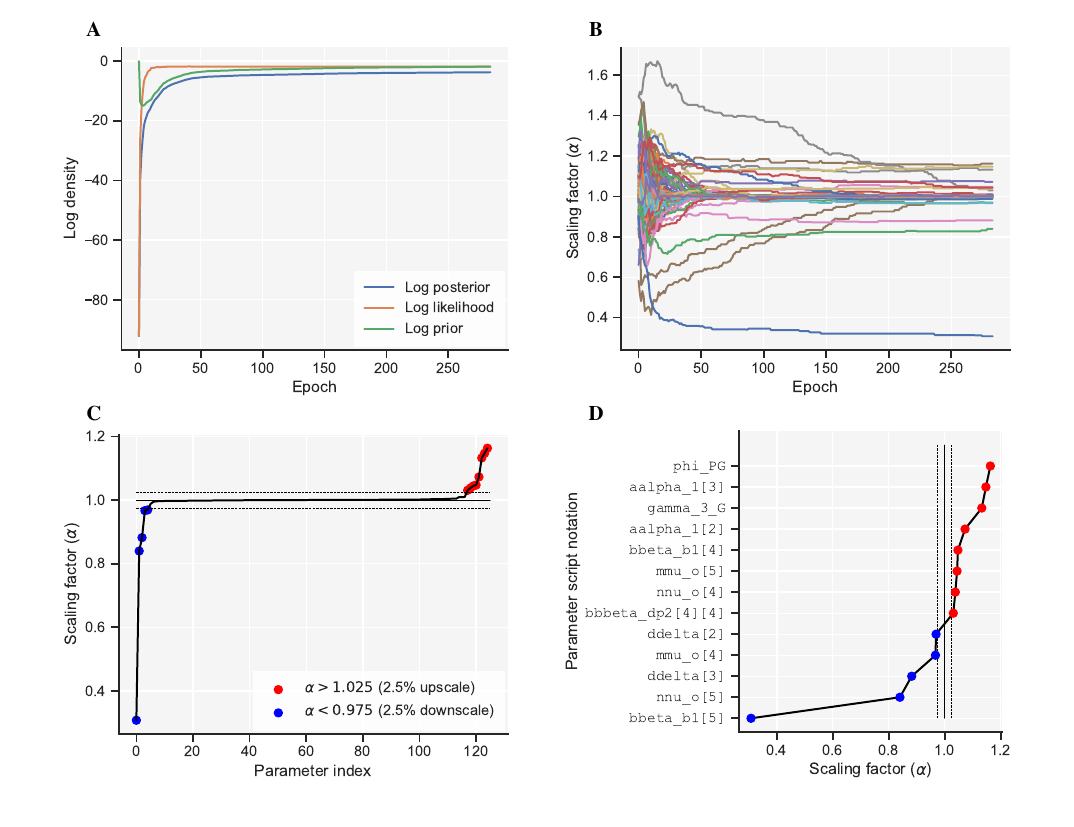}
  \caption{
    \textbf{Results of parameter optimisation.}
    Panel A shows the change in log posterior density, likelihood, and prior density with the training epoch.
    Each epoch consists of 100 steps of sparse noise differential-evolution Monte-Carlo.
    Panel B displays the trace of each element of the parameter vector, i.e. the scaling factors that upscale or downscale the prior estimates.
    Panel C shows the final value of the parameters and highlights those that correspond to a 2.5\% or more downscale (blue) or upscale (red) of the prior estimates.
    Panel D provides more details regarding the identity of the upscaled and downscaled parameters.
    In panels C--D, the upper and lower dotted lines form the 2.5\% up- and downscale range. 
    The corresponding Greek symbols are provided along with numerical values of estimates in Table S1. 
    Figure created by Willem Bonnaff\'e.
  }
  \label{fig:parameters-optimisation}
\end{figure}

\paragraph{Simulation and long-term population structure}
Long-term simulations of the best model are presented in Figure 5.
This figure shows that all variables in the ecosystem ultimately reach a stable equilibrium, which confirms our initial expectation of coexistence of the key species in this simplified version of the Yellowstone ecosystem.
The fitted model predicts asymptotic population counts with more bison than elk, 4,420 compared to 3,340 (Fig. 5, A, green and red lines), and more wolves than cougars, 100 compared to 44 (purple and brown lines).
The willow and aspen populations increase substantially, but willow density at equilibrium only reaches about half of that of aspen (Fig. 5, A, blue and orange lines).
The proportion of the total mass of organic and inorganic matter in the ecosystem accounted for by grass is comparable to nutrients but small compared to decomposers and organic matter (Fig. 5, B--C).

Our next expectation was that the model predicted plausible biomass distributions.
The mass structure of populations varies substantially throughout the time series (Fig. 5, D--I).
For example, the mass distribution of willow and aspen is dominated by larger individuals at the beginning of the time series (Fig. 5, D--E, $< 50$ years), and by smaller ones later on.
Elk and bison display a clear segregation of the population into distinct biomass groups corresponding to individuals at different stages of growth (Fig. 5, F--G).
In contrast, wolves only show two biomass groups, corresponding to the pups and adults, and cougars have a further third biomass group corresponding to larger mature individuals (Fig. 5, H--I).
Interestingly, the wolf population shows bi-annual fluctuations in population mass distribution (Fig. 5, H), with a greater abundance of smaller individuals the first year and larger individuals in the second.
Using body mass at sexual maturity to distinguish between juveniles and adults in our model, namely 150 kg for elk, 400 kg for bison, 40 kg for wolves, and 35 kg for cougars, our model predicted average body weights of adults around 245.5 kg for elk, 415.8 kg for bison, 54.9 kg for wolves, and 68.9 kg for cougars (Fig. 5, F--I.).
These predictions match relatively closely the expected mass of adult female elk (300 kg; \cite{White2024}), bison (500 kg; \cite{White2015}), wolves (40 kg; \cite{Smith2020}), and cougars (40 kg; \cite{Ruth2019}).
These quantities emerged from the simulation as there is no explicit parameter enforcing specific mean adult body mass. 
Therefore, these results provide additional evidence that the bioenergetic component of the model is capable of generating patterns congruent with the biology of the different species without explicit coding of these quantities.

\begin{figure}
  \centering  
  \includegraphics[width=\textwidth, page=5, trim={0 2cm 0 2cm}, clip]{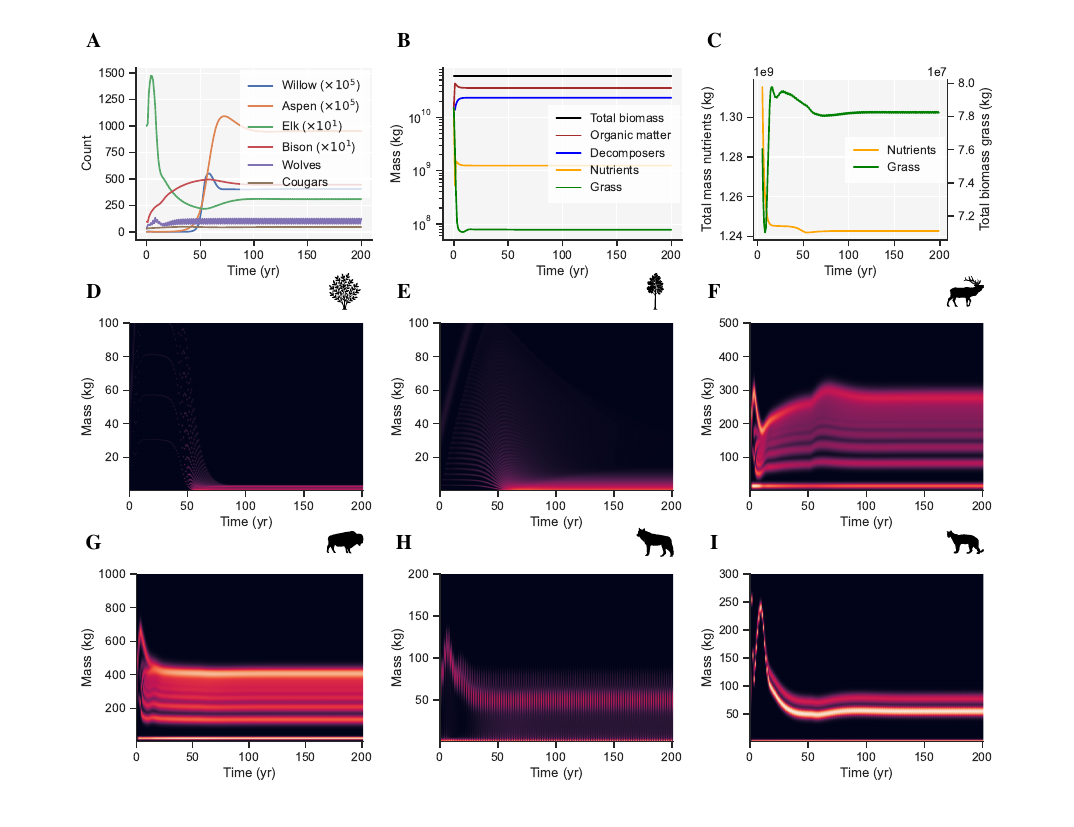}  
  \caption{
    \textbf{Simulation of the fitted Yellowstone ecosystem model.}
    Panel A shows the population counts for the structured species in the model.
    Panel B shows the total mass of the unstructured compartments and the total mass of organic and inorganic components in the system (black line).
    Panel C focusses on the mass of nutrients and grass to show finer fluctuations.
    Panel D--I show the change in the biomass distribution of the 6 structured populations through time, namely willow (D), aspen (E), elk (F), bison (G), wolves (H), and cougars (I).
    The biomass bins are obtained by discretising the biomass range of each population.
    Figure and icons created by Willem Bonnaff\'e.
  }
  \label{fig:model-simulation}
\end{figure}

\paragraph{Artificial predator extirpation and reintroduction experiment}
Our last expectation was that the model response to perturbation would match that of the real system, especially the response to predator extirpation as the model was only trained to respond to predator recovery.
Figure 6 shows the results of the artificial extirpation and reintroduction experiment.
The removal of predators after 200 years pushes the herbivores to a disrupted equilibrium point, defined by a stable population size, with a higher number of elk, about 9,200 individuals, and a lower number of bison, approximately 3,020 individuals (Fig. 6, B, brown and yellow lines).
Both aspen and willow counts decrease following predator removal.
Willow counts decrease and reach an equilibrium with fewer individuals, in the order of 20 million individuals. 
The aspen population decline is more important and prolonged than willow population decline.
The disruption times for elk and bison were about 22 years and 23 years, respectively (Fig. 6, A). 
Those were shorter than recovery times: 34 years for elk and 33 years for bison (Fig. 6, A).
Surprisingly, this pattern was reversed for willow and aspen. 
The willow population recovered its initial population size in 31 years after a disruption time of 66 years (Fig. 6, A).
Aspen reached an equilibrium within 130 years following predator removal, and recovered within 55 years of predator reintroduction (Fig. 6, A).
Increasing the time period with no predator present revealed that these equilibria were stable in the long run (Fig. S7--8).

\begin{figure}
  \centering  
  \includegraphics[width=\textwidth, page=6, trim={0 2cm 0 2cm}, clip]{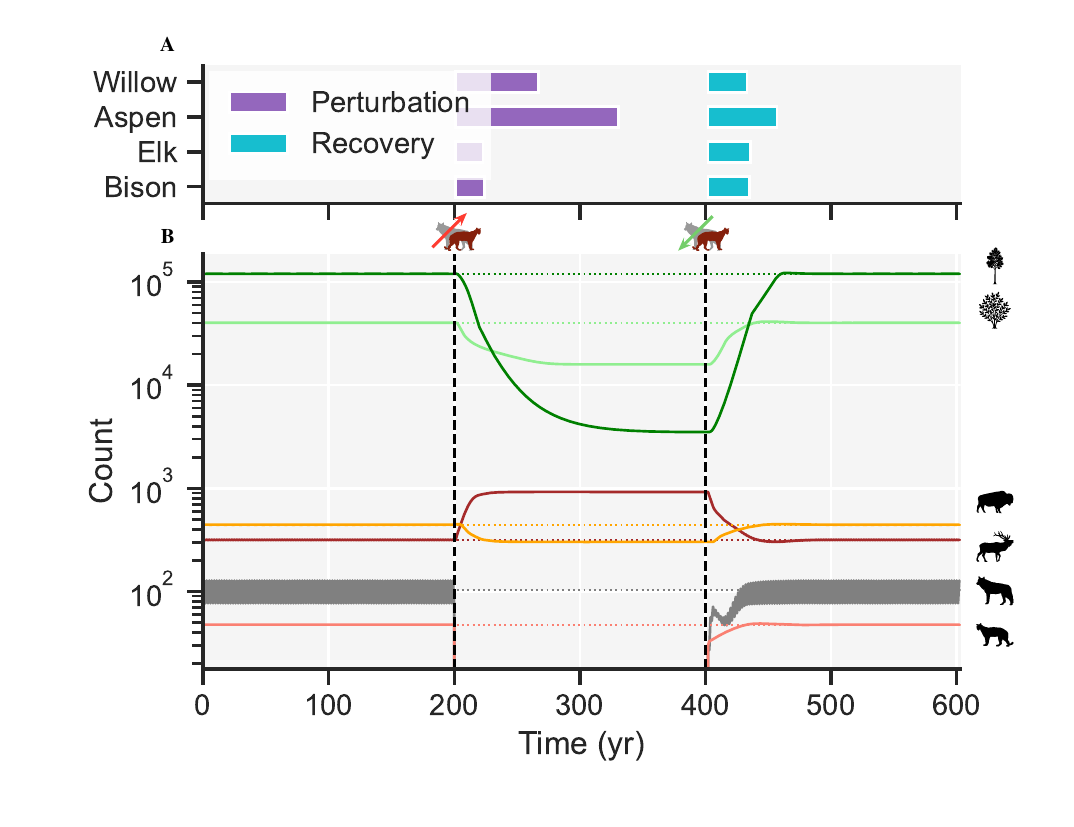}  
  \caption{
    \textbf{Artificial predator extirpation and reintroduction experiment.}
    Panel A shows the duration of the transient dynamics following the extirpation of predators (perturbation, in purple) and their subsequent reintroduction (recovery, in blue). 
    The transient part of the dynamics ends when the time-averaged number of individuals reaches an equilibrium, i.e. no temporal variations.
    Panel B displays the results of a simulation wherein the two predators (wolves and cougars) were removed from an ecosystem at equilibrium after 200 years and reintroduced 200 years later.
    Figure and icons created by Willem Bonnaff\'e.
  }
  \label{fig:extirpation-1}
\end{figure}

Figure 7 provides the detail of the mass distribution of aspen and elk following predator extirpation and reintroduction.
This reveals that the slow decline of aspen is due to older trees, immune from predation, that slowly die out, while the fast recovery is due to regeneration of stands, which will quickly produce numerous small trees (Fig. 7, C).
This figure also shows that in absence of predators, the mass distribution of elk shifts towards lower body masses, in line with stronger food limitations (Fig. 7, D).
This is confirmed by computing the elasticity of elk biomass production to resource consumption (Fig. 7, F), which shows an increased response to resources in the absence of predators.

\begin{figure}
  \centering  
  \includegraphics[width=\textwidth, page=7, trim={0 1cm 0 1cm}, clip]{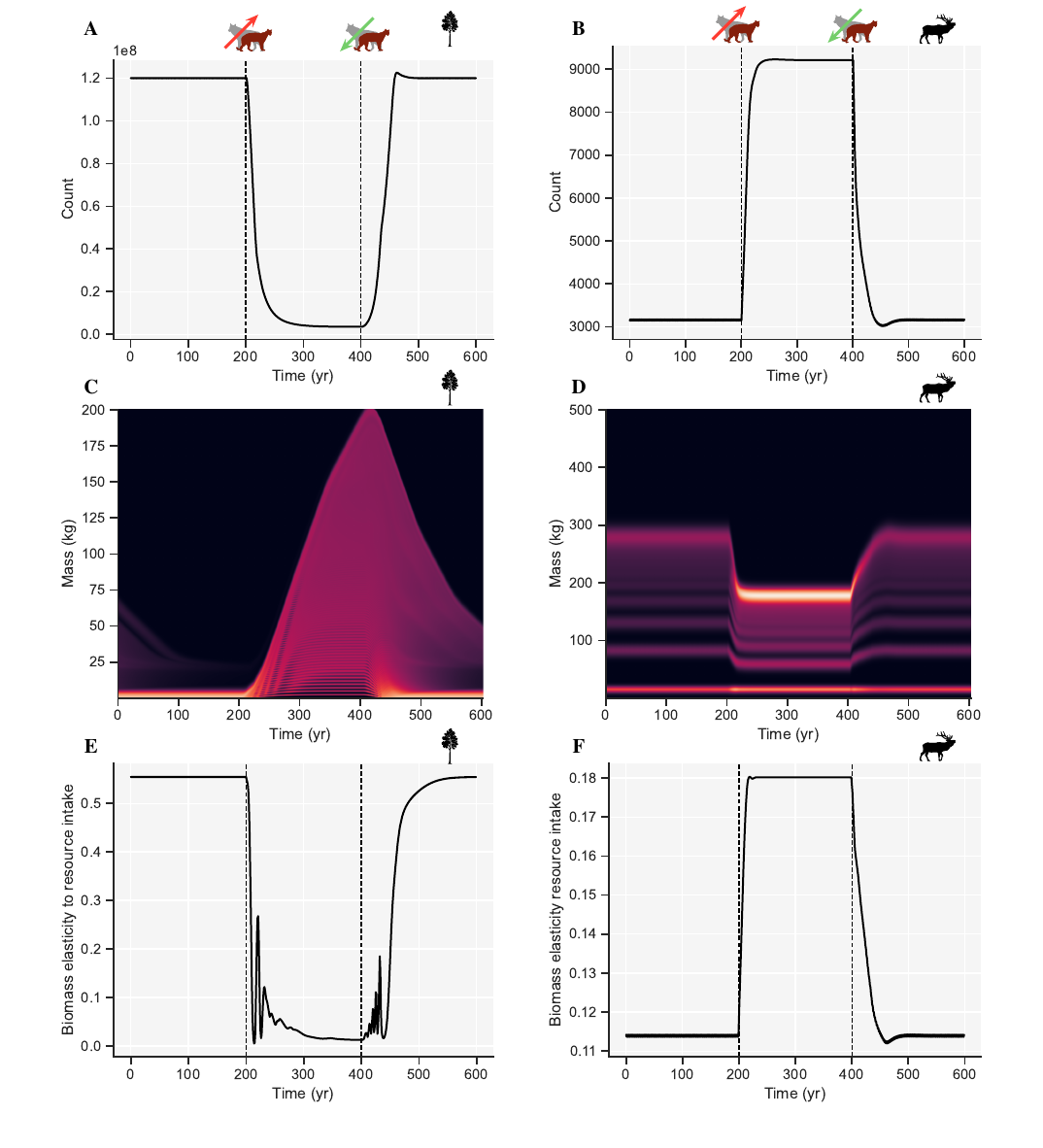}  
  \caption{
    \textbf{Changes in mass distributions as a response to predator extirpation and reintroduction.}
    Panels A and B focus on the changes in aspen and elk count, respectively, following predator removal (dashed line at the 200 year mark) and reintroduction (dashed line at the 400 year mark). 
    Panels C and D show the corresponding mass distribution at each time step. 
    Brighter colours indicate a higher density of individuals.
    Panels E and F show the elasticities of annual biomass production of aspen and elk, respectively, with respect to a change in total resource intake.
    Figure and icons created by Willem Bonnaff\'e.
  }
  \label{fig:extirpation-1b}
\end{figure}

Figure 8 provides a visualisation of the paths that the ecosystem takes on its way to the disrupted equilibrium and then back to its recovered state.
Looking at the pairwise relationships between vegetation and herbivores shows a pattern of path-dependent recovery, where the path to the disrupted ecosystem equilibrium (Fig. 8, orange dotted line and red star) is different than the path to recovery (green dotted line and green star). 
The most striking result is the relationship between elk and bison counts, which reveals that the removal of predators and subsequent reintroduction leads to a transition from a predator-free ecosystem state where bison is relatively less abundant compared to elk (red star), to a state where it is relatively more abundant (green star). 
This imbalance is not found for vegetation; looking at willow and aspen shows that both aspen and willow counts are relatively high, or low, when predators are present (green star), or absent (red star), respectively, though the path to recovery is characterised by a relatively higher initial density of willow compared to aspen (green dotted line).
Another marked pattern of path-dependent recovery can be observed by considering the relationship between aspen and elk, where the recovering system (green dotted line) goes through a period with relatively fewer aspen and higher elk counts, followed by a period of higher aspen and lower elk, whereas the disrupted system follows a trajectory where both elk and aspen overlap at intermediate densities (orange dotted line). 

\begin{figure}
  \centering  
  \includegraphics[width=\textwidth, page=8, trim={0 2cm 0 2cm}, clip]{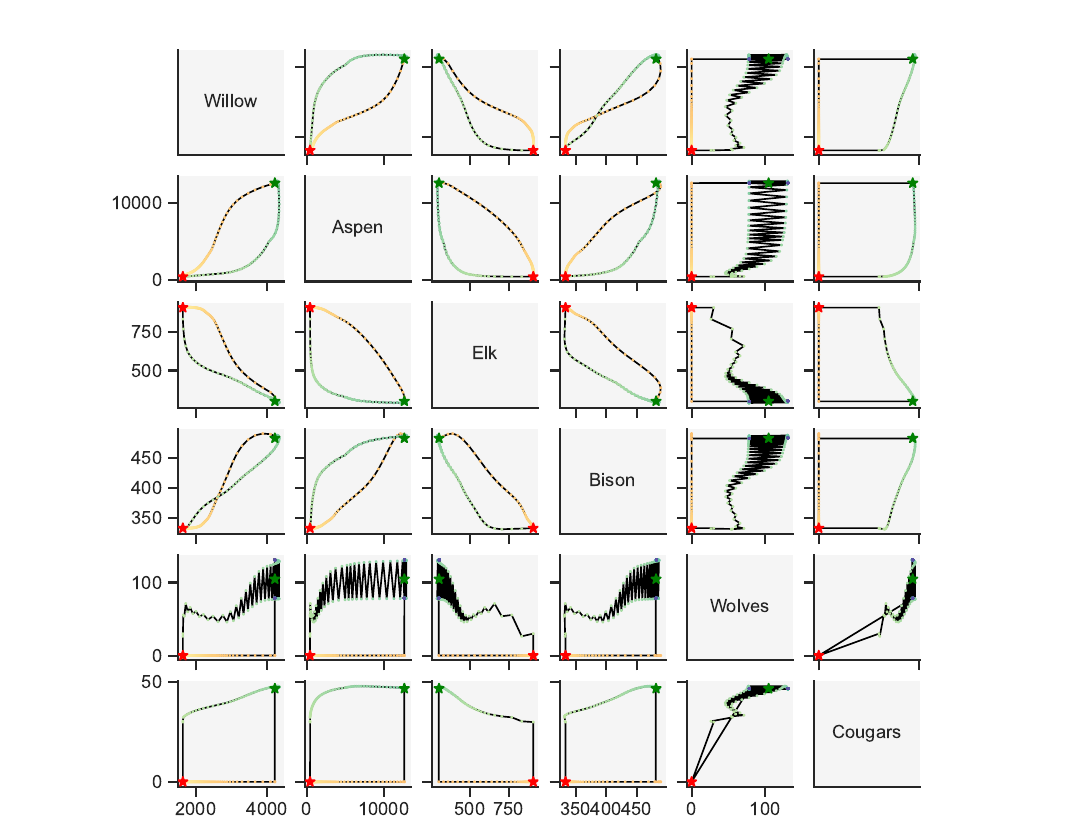}
  \caption{
    \textbf{Consequences of predator extirpation and reintroduction on ecosystem states.}
    The panels show the path of the ecosystem between two equilibria, considering pairwise combinations of states. 
    The disruption path (yellow dotted line), immediately following predator extirpation, leads to the disrupted equilibrium (red star).
    The recovery path (green dotted line), which follows the reintroduction of the predators, leads back to the initial equilibrium (green star).
    Figure created by Willem Bonnaff\'e.
  }
  \label{fig:extirpation-2}
\end{figure}

\paragraph{Consistency of results across models}
We assess here the consistency of the results presented in the previous sections across our different models.
We consider the second-best model fitted to the entire time series, and the best model fitted only to the first 20 years.
The most up- or down-scaled parameters can be compared across models by looking at Figure 4 and Figure S2.
We find that the parameter scalings are broadly similar, i.e. $\phi_{PG}$, $\alpha_1^{(4)}$, $\gamma_3$, $\mu^{(6)}_{o}$ are upscaled by more than 2.5\%, while $\delta^{(3)}$, $\delta^{(4)}$, $\nu_o^{(6)}$, $\beta_{b,1}^{(6)}$, $\mu_o^{(5)}$ are downscaled by more than 2.5\%, in all three models.
$\beta_{b,1}^{(5)}$ is downscaled both in the best and second-best model but not in the model trained on the first 20 years.
$\beta_{d,p,2}^{(5,5)}$ is upscaled only in the best model.
The response of the second-best model to predator extirpation is presented in Figures S3--4 while that of the model trained only on 20 years is shown in Figures S5--6.
These figures show that the response to predator extirpation/recovery is virtually identical across models.
The main results are conserved, i.e. that the woody deciduous vegetation and bison population decline following predator removal, and recover following predator recovery (Fig. S3 and S5).
This pattern is underpinned by shifts in biomass distributions and growth limitations governed by the presence or absence of predators (Fig. S4 and S6).

\section*{\uppercase{Discussion}}

Our aim in this work is to further our understanding of how individual-level processes, and in particular survival, growth, and reproduction, link to ecosystem-level processes, through structured interactions within and across species. 
To address this, we introduce a novel framework, bioenergetic ecosystem integral projection modelling (Eco-IPM).
This work advances mathematical models of continuously-structured populations in four ways. 
First, we integrate ecosystem-level processes with individual-level processes, through the conversion of biomass fluxes into bioenergetic allocation to individual growth, reproduction, and maintenance. 
Second, we include multiple structured species in multiple different trophic groups, thus including all essential combinations of competition and predatory interactions in food webs. 
Third, we introduce nutrient recycling, which is a key component of ecosystem functioning. 
Finally, we develop an optimisation algorithm suitable for fitting IPMs to time series of count data, thus solving the key challenge posed by the parameterisation of large models and grounding within real systems. 
We show how our approach can be used to study ecosystem dynamics in northern Yellowstone, focusing on the role of predators in explaining the changes in herbivore-vegetation states that have been observed over the past century. 
We discuss below these technical and biological contributions in light of the literature.

\subsection*{Linking individual bioenergetics to ecosystem dynamics}
The first advance consists of a mechanistic link between the life history of individuals and biomass fluxes at the level of the ecosystem. 
This is achieved by combining two steps. 
First, we used a binomial model for food capture, following which the biomass distribution of each population is split according to two different sources of mortality, either intrinsic or due to consumers, and allocated to the different consumer populations responsible. 
Second, we partitioned the biomass consumed into maintenance, reproduction, or growth. 
These steps have been introduced independently in previous work (\cite{Bassar2023, Passoni2024}). 
Bassar et al. (2023) split the survival of guppies and killifish into two components, one of which represents survival to predation by the other species. 
However, this work considered body length as the phenotypic trait, rather than body mass, and so does not enable tracking biomass fluxes among species. 
The second study, by Passoni et al. (2024), used the same bioenergetic model as in our study, but considered that wolf density reduces the log-odds of survival of elk, which does not allow for a decoupling of intrinsic mortality from consumer-related sources of mortality. 
Therefore, this construction could not be used for allocating biomass to the different consumer populations, and therefore for tracking ecosystem biomass fluxes. 
Our study hence provides a mechanistic link between the biomass fluxes between and within species.

The second and third advances are (1) that we provide a general model that enables the inclusion of structured populations at every trophic level of an ecosystem, from primary producers to predators, and (2) that we embed this model in a nutrient recycling loop including decomposers. 
Our framework thus enables the study of multiple types of interactions in a minimal but complete ecosystem context, while other studies to date tended to tackle those in isolation. 
The inclusion of body mass structure introduces implicitly indirect intra-specific competition for food capture, as opposed to direct competition where individuals actively contest or fight for resources. 
The inclusion of multiple species at each trophic level introduces indirect inter-specific competition. 
The inclusion of multiple trophic levels introduces structured interactions between primary producers and primary consumers, and between primary consumers and secondary consumers. 
The predator populations (secondary consumers) are also subject to intra-specific predation. 
While we did not consider intraguild predation in our case study of northern Yellowstone, due to lack of consistent evidence for direct interactions between wolves and cougars (\cite{Smith2020, Rabe2025}), this type of interaction can be included in our framework but its strength is set to zero. 
Importantly, the inclusion of three trophic levels and nutrient recycling allows for indirect interactions between primary producers and predators through effects on primary consumers and decomposers, which so far has not been achieved in similar work. 
Bassar et al. (2023) achieved one of the most exhaustive integrations of different structured interaction types within a single system to date, by including intra- and inter-specific competition, intra-specific predation, and intraguild predation, in a two-species IPM. 
However, this model included structure at a single trophic level. 
Therefore, by extending this to multiple other trophic levels, our work creates novel avenues for investigating the relationships between interactions of different types in food webs and their consequences for ecosystem dynamics.

The last advance that we introduce in this paper is technical. 
We provide an optimisation algorithm that automatically calibrates the parameters of the model based on time series data of animal counts, thus grounding the model in a real system. 
This step is usually the bottleneck in most modelling work and can lead to arbitrary parameter choices. 
While our methodology still relies on a thorough literature search of initial parameter values and validation by experts, the automatic calibration removes some of the subjectivity of this initial step. 
It further relies on an innovative combination of difference equations (\cite{TerBraak2006}), a highly flexible global optimisation algorithm requiring minimal tuning (\cite{Bonnaffe2022}), and Bayesian regularisation (\cite{Cawley2007}), which enforces parsimonious parameter values. 
Previous studies have laid a solid foundation for performing Bayesian inference of parameters in IPMs using Markov chain Monte Carlo (\cite{Elderd2016, White2016}). 
Yet, the work so far has relied on data on the relationship between phenotypic traits and individual-level processes, e.g. survival, growth, and reproduction, to estimate parameters in single population models. 
Individual-level data is labour-intensive to collect and is thus not available for most systems, especially when multiple species are involved. 
Our optimisation approach hence partly relaxes this constraint by only requiring population-level data. 
Grounding the model means that, beyond its usefulness for theoretical investigations, our framework also constitutes a tool to make inferences and predictions regarding the real systems that it is based on. 
This can be useful to test the effect of different population control strategies, or to determine the state of the system in the future, or in the past, before any data was collected.

\subsection*{Application to northern Yellowstone}

We use our framework to study the impact of predator extirpation and reintroduction in northern Yellowstone. 
We calibrate a model featuring three key primary producers (grass, willow, and aspen), two primary consumers (elk and bison), and two predators (wolves and cougars) using time series collected in northern Yellowstone. 
We are confident in the inferences drawn from our Yellowstone ecosystem model because it produces plausible biological dynamics and quantities that were not directly optimised for. 
First, we found agreement between average adult body masses derived from the literature and those predicted by the model, though these quantities are not explicitly enforced by specific parameters.
Second, the response of the model ecosystem to predator extirpation is congruent with the history of Yellowstone, though the model was only trained to respond to predator reintroduction, and not extirpation.
Finally, we performed validation and testing on unseen data, which showed that the model is capable of generalising to new time steps, and that all inferences of the model were robust even when part of the time series were withheld at the training stage.
Overall, this confirms that bioenergetics can produce plausible quantities and dynamics even when they are not explicitly enforced by the model structure (\cite{Smallegange2017}), and that this property holds in the more complicated model considered here.

Aside from deviations in parameter values and literature estimates, we also observed deviations of model predictions from the empirical observations, especially in the aspen and willow populations. 
Our model predicts an aspen population around a hundred million individual aspen trees across all sizes, as identified by tree stems rather than by genotype, for the entire northern Yellowstone area. 
In reality, these numbers are likely to be much smaller; perhaps closer to fifty thousand individuals.
A potential cause for this discrepancy is that our model does not account for climate change, which is predicted to decrease aspen abundance in northern Yellowstone (\cite{Piekielek2015, Piekielek2016}), along with droughts and infrequent fires in the area (\cite{Krasnow2015, Morris2019}). 
Our model predicts that, in the absence of adverse climate-change effects, the northern Yellowstone aspen population may remain viable, suggesting that predator-mediated reductions in herbivory could contribute to its persistence within our simplified representation of the ecosystem. This model-based inference should be interpreted cautiously, however, because empirical evidence for broad recovery of aspen and willow remains mixed (\cite{Hobbs2024, Painter2025, Beschta2026, Hobbs2026,  MacNulty2026}).
Future implementations could extend the abiotic components in the model, for example by including water limitations to account for hydrology and light limitations on plant growth, to quantify the relative contribution of abiotic factors compared to biotic factors to ecosystem dynamics.

Our simulation experiment of predator extirpation and reintroduction predicted ecosystem-wide consequences of predator removal. 
First, we found that effects of predator removal were fully reversible, though recovery took more time than disruption and involved a different path in the ecosystem space, providing further evidence for path-dependent recovery in Northern Yellowstone (\cite{Hobbs2024}). 
Second, we identified a counter-intuitive response of herbivores. 
While the elk population in our model ultimately increased following predator extirpation, consistent with the post-1969 increase of elk in northern Yellowstone (\cite{White2024}), the bison population decreased and only recovered after predator reintroduction. 
This suggests that wolf reintroduction, through its negative impact on elk, potentially contributed to the recent increase of bison in northern Yellowstone, which is an intriguing hypothesis that has been proposed but not tested to date (\cite{Ripple2010}). 
We attribute this pattern to the fact that elk, being a browser, has access to additional resources and is therefore able to grow to a larger population size than if it was relying on grass alone.
This in turn reduces the grass biomass levels, and consequently, increases resource limitations experienced by the bison population.
In reality, browsing also occurs in bison, but leaves of woody plants comprise less than 5\% of their diet (\cite{White2015}), compared to 10--30\% in elk (\cite{White2024}), which is why we did not account for browsing in bison in our model.
Furthermore, through tracking species biomass distributions and estimating resource limitations, we showed that population changes continue to occur even after the counts have reached a stable equilibrium.
Predators contribute to relaxing resource limitations on population growth.
This is another way in which bioenergetic IPMs, by tracking biomass distributions and resource intake, can provide more mechanistic insights into population declines or increases.
Apart from herbivores, predator extirpation and reintroduction also impacted woody deciduous vegetation. 
We observed a higher density of individual aspen trees and willow shrubs in presence of predators, than when they are absent.
This is also consistent with the signs of regeneration detected in localised stands of aspen and willow in northern Yellowstone following the reintroduction of wolves in 1995 (\cite{Hobbs2024}), although evidence of regeneration on a wider scale is limited (\cite{MacNulty2025, MacNulty2026}).
Other studies predict declining habitat suitability for aspen in the study area due to climate warming, rather than due to browsing by herbivores (\cite{Piekielek2015, Piekielek2016}).
Overall, our results support the important role that predators play in structuring ecosystems, though detecting clear patterns of ecosystem response to predator restoration in the field are often obscured by methodological, conceptual, and biological complications (\cite{Brice2022, Hobbs2024, Beschta2026, Hobbs2026}). 

\subsection*{Complicated models with simple dynamics?}

Our results also yield theoretical insights, first by revealing a fixed-point equilibrium with coexistence of all species. 
This contrasts with pioneering work in theoretical ecology that showed that chaos could arise even in a single population (\cite{May1976}), setting the expectation for large systems to be unstable. 
In spite of the relative complexity of the model, we did not find evidence of multi-stability or intricate multi-dimensional oscillations, similar to how simpler linear models behave.
Our work hence provides an additional example of a system where numerous structured interactions can lead to stability and resilience to perturbations, at least in the minimalistic ecosystem configuration considered here compared to the complexity of real ecosystems (\cite{McCann2000}).
In other words, complicated models can also produce simple dynamics.
In spite of this apparent simplicity of dynamics at the community level, our model also produced unanticipated dynamical patterns within populations.
For example, our model produced fluctuations in wolf count that resemble those in the time series through bi-annual cycles in the mass structure of the population, whereby the population is dominated by sub-adults in one year, and by larger mature individuals the following year.
This pattern is not observed in the real system, at least not in this clear form.
We attribute this to the fact that wolves generally require two years of growth before they reach sexual maturity, for this reason it is possible that in some conditions the population enters an age structure cycle, especially when survival is low.
This pattern may also be due to intra-specific wolf mortality linked to conflicts between packs, which reduces survival further at higher densities (\cite{Smith2024}).
Overall, these examples only scratch the surface of the great variety of dynamical patterns that more complex models can generate.
Further efforts to characterise those and dissect them mechanistically will follow.

\subsection*{Limitations and prospects}

While we believe that our work constitutes a key step towards more realistic models of ecosystem dynamics, there are several areas that warrant further efforts. 
The ecosystem configuration considered here is deliberately minimalistic to facilitate further model extensions, and so it does not include multiple factors that are key to understanding ecosystem dynamics. 
Senescence, which can lead to a decrease in reproductive output and survival in later stages of life, is not accounted for.
This could be done by introducing age-structure into the model.
In addition, spatial dynamics are not considered, though they are often an essential mitigating factor of interactions between individuals, especially in northern Yellowstone where both elk and bison undergo seasonal migrations (\cite{Geremia2019, White2024}). 
Parasites and disease are not present, but they also play a key role in regulating populations (\cite{Hudson2006}). 
For instance, canine distemper virus (CDV) outbreaks and brucellosis are a substantial source of mortality in predators and herbivores across multiple systems, including northern Yellowstone (\cite{White2015, Smith2020}). 
Pack dynamics, namely fluctuations in the social structure of wolf packs, also have bearings on the capacity of a pack to maintain its territory and acquire food (\cite{Smith2020}), and therefore on the dynamics of other species in the ecosystem.
Introducing context dependence in interactions between species and their environment, for example by implementing temperature dependence in consumption rates, and metabolic costs, or disease-mediated predation, would be a fascinating way to explore how varying orders of complexity in interactions affect ecosystem dynamics (\cite{Kleinhesselink2022}). 
Emerging methodologies combining neural networks and dynamical systems may help in identifying the level of complexity of these interactions (\cite{Bonnaffe2021, Bonnaffe2023, Zappala2024}).
Finally, abiotic factors, such as droughts, wildfires, and winter severity, also drive transitions in ecosystem state and successions (\cite{Smith2020, White2024}). 
These last drivers are magnified by climate change making it a pressing area of research. 
Our framework provides an ideal foundation for studying the impact of these stressors on ecosystem dynamics.
We propose a simple conceptual framework to guide these investigations in Figure 9.

\begin{figure}
  \centering  
  \includegraphics[width=\textwidth, page=9, trim={0 1cm 0 1cm}, clip]{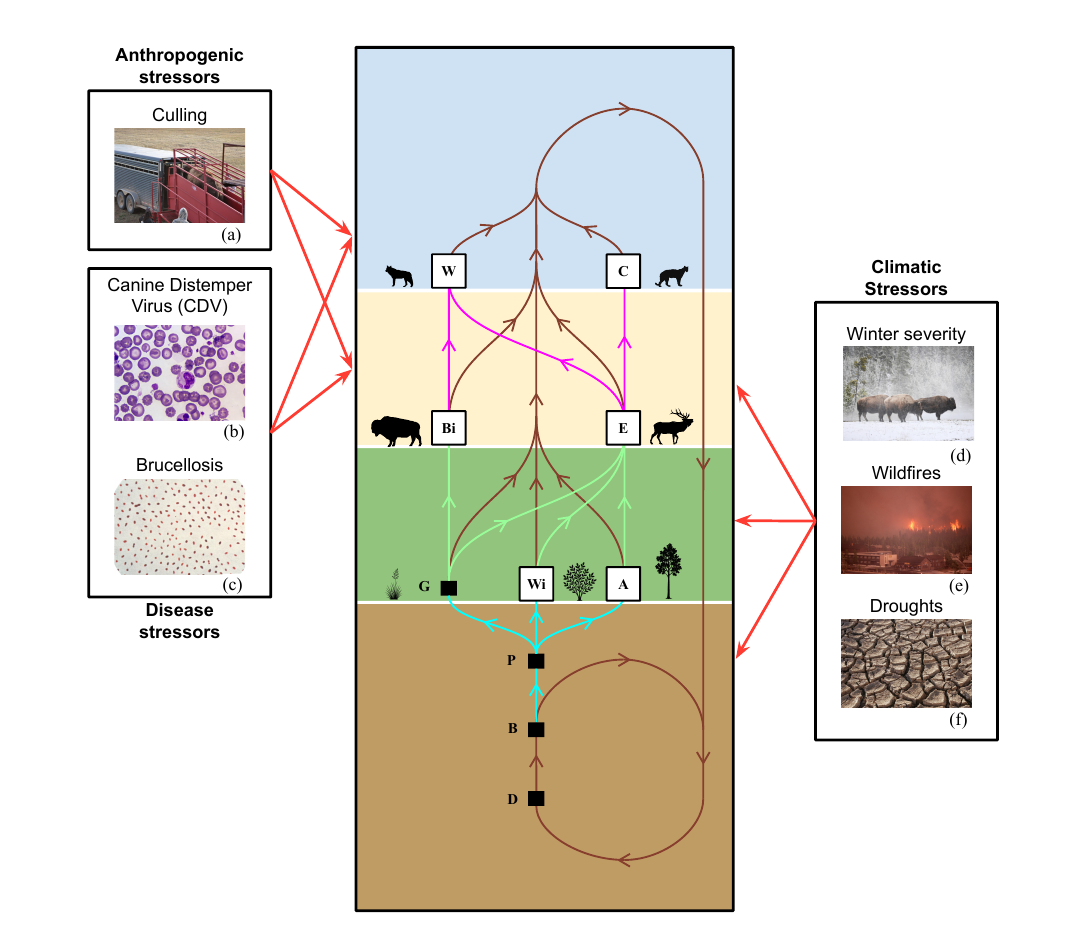}
  \caption{
    \textbf{Potential points of impact of anthropogenic, disease, and climatic stressors on ecosystems.}
    Thick curvilinear lines represent biomass fluxes between the different components of the ecosystem.
    Straight red arrows correspond to direct impacts on the main ecosystem compartments, namely soil (brown), primary producers (green), primary consumers (yellow), and secondary (and higher order) consumers (blue).
    The unstructured compartments of the ecosystem are decomposing organic matter (D), decomposers (B), nutrients (P), and grasses (G).
    The structured compartments are willow shrubs (Wi), aspen trees (A), bison (Bi), elk (E), wolves (W), and cougars (C).
    Culling, which corresponds to the removal of individuals from various populations, affects primary and secondary consumers.
    Canine distemper virus (CDV) and brucellosis are typical examples of disease stressors. 
    CDV affects predators, while Brucellosis affects herbivores. 
    Climatic stressors can directly affect the soil, primary producers, and primary consumers.
    Figure and icons created by Willem Bonnaff\'e.
    Photos licence and credits: 
    (a) Public domain (\href{https://commons.wikimedia.org/wiki/File:Culled\_bison\_are\_transported\_to\_their\_new\_homes\_on\_tribal\_lands\_and\_in\_other\_parks\_and\_preserves\_on\_special\_trucks.\_\%288233a5f0-1dd8-b71b-0b51-50ab26c903e1\%29.JPG}{U.S. National Park Service}), 
    (b) \href{https://creativecommons.org/licenses/by/4.0/}{CC BY 4.0} (\href{https://commons.wikimedia.org/wiki/File:Canine\_Distemper\_Virus\_Cytoplasmic\_Inclusion\_Body\_\%28Blood\_smear,\_Wright\%27s\_stain\%29.jpg}{Lance Wheeler}), 
    (c) Public domain (\href{https://commons.wikimedia.org/wiki/File:Brucella\_melitensis.jpg}{Centers for Disease Control and Prevention}),
    (d) Public domain (\href{https://commons.wikimedia.org/wiki/File:Bison\_in\_Snow\_-\_Flickr\_-\_YellowstoneNPS.jpg}{Kevin C. Weber}), 
    (e) Public domain (\href{https://commons.wikimedia.org/wiki/File:Fire\_near\_Old\_Faithful\_Complex\_2.jpg}{Jeff Henry}), and
    (f) \href{https://creativecommons.org/licenses/by/3.0/}{CC BY 3.0} (\href{https://commons.wikimedia.org/wiki/File:Drought.jpg}{Tomas Castelazo}).
  }
  \label{fig:ecosystem-stressors}
\end{figure}

Apart from these aspects, which are simply not present in the model, there are a number of processes that we included that are simplistic representations of the real processes. 
First, while it is a step in the right direction, including nutrient cycling as a system of simple difference equations is a drastic simplification which flattens the diversity of the agents that decompose organic matter and of their interactions.
Even at the functional group level, we have not accounted for microbivores, which can introduce density-dependent feedback within the decomposer community dynamics.
Possibly as a consequence of this simplification, our model predicts decomposer and organic matter masses that are so large that the influence of animals and deciduous plants in our model ecosystem on its dynamics are relatively inconsequential.
Future work could introduce structure in these compartments as a way to emulate their diversity of functions and study its consequences for ecosystem functioning.
Additionally, we only implemented indirect competition between consumers that rely on the same food sources.
This is a simplification given that, for example in northern Yellowstone, willow and aspen exist in different types of habitats, as willow grows in riparian areas (\cite{Hobbs2024}), whereas aspen is primarily an upland species. 
These two species hence tap into pools of resources that are to some extent inaccessible by the other.
This could be accounted for by modelling explicitly ecosystem compartments.
Finally, some species in the model affect their ecosystems in more complex ways than considered here due to their ecosystem engineer role (\cite{Sanders2024}).
It is the case of bison which can alter the rate of biomass production in grass by increasing soil nutrient content (\cite{Geremia2019}).
Other examples in Yellowstone are beavers that raise the water table by building dams, thereby expanding the riparian area and altering the competition between willow and aspen (\cite{Hobbs2024}).
These ecosystem engineers require specific model implementations that will be the object of dedicated follow-up studies.

A more technical limitation of our approach is that it requires literature-informed initial parameter values and longitudinal data. 
It is possible that certain combinations of initial parameters lead to unrealistic population dynamics, extinctions, or simply prevent computations due to numerical errors, though we did not find this to be the case for the minimal northern Yellowstone ecosystem considered here. 
This is why our framework still requires a thorough literature search for reasonable parameter values and validation by experts. 
It also requires longitudinal data for calibration, which for many systems and species is not available. 
This was the case in our study for willow and aspen biomass.
Although estimates exist for selected locations, they were not sufficient to constrain the predictions of the model for the whole Northern Range.
Future implementation of this framework would need to find a way to constrain the dynamics of trees and shrubs, potentially by using predictions of larger more mechanistic models where data is absent.
Altogether, the application of our framework to less well-studied ecosystems is likely to be limited by the availability of system-specific knowledge. 
This limit may be partly lifted with the rise of high quality biological databases and machine learning methods for dealing with partial knowledge (e.g. Neural ODEs, \cite{Bonnaffe2021, Bonnaffe2023}). 

This nonetheless leaves exciting prospects whereby our framework can be applied to well-studied systems, for instance to Kruger National Park (South Africa), or the Windermere system (United Kingdom) for studying freshwater ecosystems, where extensive data on individual processes of growth, survival and reproduction, as well as time series of population counts are available, which could lead to applications to park management and fisheries.
Our current implementation of the model for northern Yellowstone can be translated with minimal changes to model a savannah ecosystem, as both ecosystems feature species that fulfill similar ecological functions. 
For instance, predators could be lions (\textit{Panthera leo}) and leopards (\textit{Panthera pardus}), large herbivores the greater kudu (\textit{Tragelaphus strepsiceros}) and African buffalo (\textit{Syncerus caffer}), and various species of protected vegetation greenthorn (\textit{Balanites maughamii}) and false tamboti (\textit{Cleistanthus schlechteri}).
This would require adjusting the parameters of the present species so that they reflect better the biology of these African species.

Our main aim in this manuscript was to compare our model predictions to a relatively well-characterised ecosystem response. 
This was to assess whether the model behaves according to our expectations. 
Future studies building upon this work will answer biological questions that are less well understood. 
For example, the study of the biomass fluxes between species predicted by the model can help us quantify the relative importance of top-down and bottom-up regulation in Yellowstone. 
Additionally, tracking biomass distributions also enables the assessment of the impact of biomass offtakes by hunters outside of the park on the Yellowstone trophic cascade.
Other questions that could be studied with this framework range from population management, i.e. determining the potential impact of different culling levels in the different species, to quantifying the interaction between the species to focus conservation efforts on species with the greatest ecosystem impacts, through to studying the impact of disease and poaching on structured populations.
These last applications would require addressing the limitations in the vegetation compartments discussed previously and a more extensive test of the model's capacity to generalise to new time steps before predictions could be put into practice, and thus, ideally, longer time series than the ones considered here.
In general, our model provides a framework to determine the role of the various species, and in particular of predators, in driving the dynamics of entire ecosystems, and how these relationships may be altered by anthropogenic stressors, disease, and climatic stressors (Fig. 9).


\paragraph{Acknowledgments}
This work was supported by UK Research and Innovation through the Horizon Europe Guarantee [Grant number EP/Y029720/1].
We thank Dean Pearson, Peter Adler, and an anonymous reviewer for their constructive criticism and feedback.
We thank Stella M. Felsinger for insightful feedback throughout the development of this work.

\paragraph{Authors' contributions}
\textbf{Willem Bonnaff\'e:} Conceptualisation, Methodology, Software, Validation, Formal analysis, Data Curation, Writing - Original Draft, Visualisation. 
\textbf{Martina Muraro:} Conceptualisation, Writing - Review \& editing. 
\textbf{William Goulding:} Writing - Review \& editing. 
\textbf{Doug W. Smith:} Writing - Review \& editing. 
\textbf{Dan R. Stahler:} Resources, Data Curation, Writing - Review \& editing. 
\textbf{Peter Hudson:} Conceptualisation, Writing - Review \& editing. 
\textbf{Hamish McCallum:} Conceptualisation, Writing - Review \& editing. 
\textbf{Sonya Clegg:} Conceptualisation, Writing - Review \& editing. 
\textbf{Dan R. MacNulty:} Conceptualisation, Methodology, Investigation, Resources, Data Curation, Writing - Review \& editing, Supervision. 
\textbf{Tim Coulson:} Conceptualisation, Methodology, Investigation, Resources, Writing - Review \& editing, Supervision, Project administration, Funding acquisition. 

\paragraph{Conflict of interest}
None of the authors of this manuscript have a conflict of interest to declare.

\paragraph{Open research statement}
All code, both for the curation and preparation of the time series and for the definition and analysis of the model, is provided in a GitHub repository (\url{www.github.com/willembonnaffe/Eco-IPM}) and archived along with the dataset.
The code can be executed online using Google Colab.
First, in Colab (\url{https://colab.research.google.com/}), open a notebook using the GitHub option.
Then enter the URL of the Eco-IPM GitHub repository (\url{https://github.com/WillemBonnaffe/Eco-IPM}). 
Select a notebook and follow the instructions in the notebooks to run the code online.

\end{spacing}
\printbibliography

\clearpage
\includepdf[pages=-]{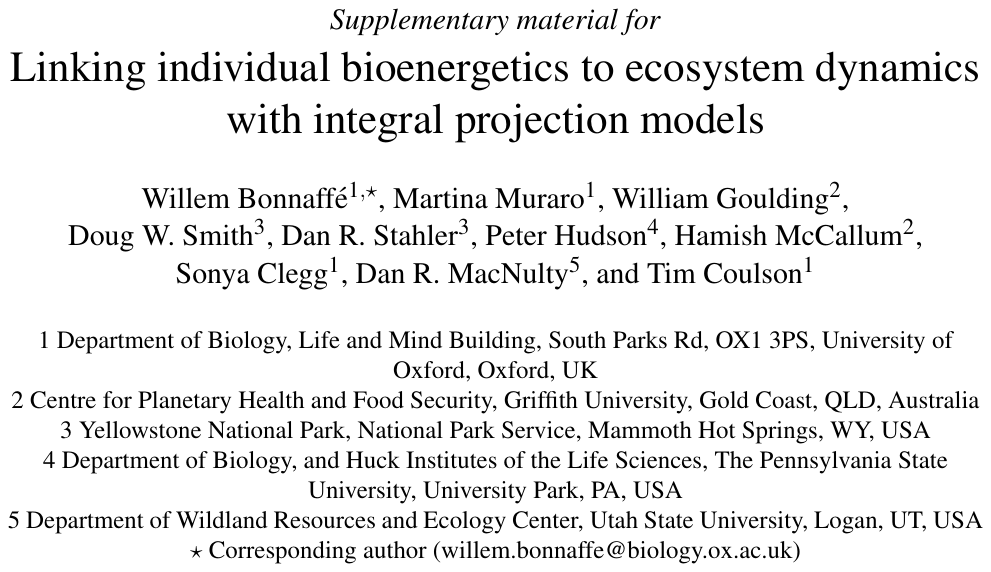}

%
\end{document}